\documentclass[11pt,article,nofootinbib,notitlepage]{revtex4-1}
\usepackage{amsmath,amssymb,float,color}
\usepackage[pdftex]{graphicx}

\def\bp{{\boldsymbol p}}
\def\bJ{{\boldsymbol J}}
 
\def\pperp{p_{\!\perp}}
\def\ttt{2$\leftrightarrow$2\ }

\newcommand{\ltsim}{\protect\raisebox{-0.5ex}{$\:\stackrel{\textstyle <}{\sim}\:$}}
\newcommand{\gtsim}{\protect\raisebox{-0.5ex}{$\:\stackrel{\textstyle >}{\sim}\:$}}

\begin{document}

\title{Effective kinetic description of the expanding overoccupied Glasma}
\author{Naoto Tanji}
\affiliation{Institut f\"{u}r Theoretische Physik, Universit\"{a}t Heidelberg, Philosophenweg 12, 69120 Heidelberg, Germany}
\author{Raju Venugopalan}
\affiliation{Physics Department, Brookhaven National Laboratory, Bldg. 510A, Upton, NY 11973, USA}
\date{\today}

\begin{abstract}
We report on a numerical study of the Boltzmann equation including $2\leftrightarrow 2$ scatterings of gluons and quarks in an overoccupied Glasma undergoing longitudinal expansion. We find that when a cascade of gluon number to the infrared occurs, corresponding to an infrared enhancement analogous to a transient Bose-Einstein condensate, gluon distributions qualitatively reproduce the results of classical-statistical simulations for the expanding Glasma. These include key features of the distributions that are not anticipated in the ``bottom-up" thermalization scenario. We also find that 
quark distributions, like those of gluons, satisfy self-similar scaling distributions in the overoccupied Glasma. We discuss the implications of these results for a deeper understanding of the self-similarity and universality of parton distributions in the Glasma. 
\end{abstract}

\maketitle

\section{Introduction} \label{sec:intro}

There are several remarkable and puzzling features of the early time nonequilibrium evolution of quark-gluon matter in ultrarelativistic heavy-ion collisions. A systematic study of its properties is feasible in the limit where the QCD coupling is very weak and the occupancies of gluons are correspondingly large such that their product $g^2\,f_g\sim \mathcal{O}(1)$. In this limit, the spacetime evolution of the overoccupied matter--the Glasma~\cite{Kovner:1995ja,Lappi:2006fp}--can be studied employing real-time classical-statistical simulations~\cite{Krasnitz:1998ns,Lappi:2003bi} of the Yang-Mills equations. Recent large scale 3+1-dimensional Yang-Mills simulations of the expanding Glasma clearly demonstrate that it flows to a nonthermal turbulent fixed point characterized by a self-similar gluon distribution~\cite{Berges:2013eia,Berges:2013fga}. Rather unexpectedly, the simulations showed that key properties of  the gluon distribution were identical to those anticipated in the ``bottom-up" thermalization scenario of Baier et al.~\cite{Baier:2000sb}. 

The results of the classical-statistical simulations are unexpected because gluon dynamics in the expanding Glasma shows no trace of late time plasma instabilities--these should be present in weak coupling kinetic frameworks~\cite{Mrowczynski:1993qm,Kurkela:2011ti,Mrowczynski:2016etf}. Further, the ratio of the longitudinal pressure to the transverse pressure $P_L/P_T$ deviates from the expectation from kinetic theory--including as well, that from the bottom-up scenario. Finally, and most remarkably, the properties of the nonthermal Glasma fixed point are, in a wide inertial range of momenta, identical to those of self-interacting scalar fields prepared with the same geometry~\cite{Berges:2014bba}. Put together, these observations pose a challenge for straightforward kinetic descriptions of the expanding Glasma~\cite{Berges:2015ixa}. 

In this paper, we will investigate the nonequilibrium early time evolution of quark and gluon matter in the longitudinally expanding geometry of an ultrarelativistic heavy-ion collision by numerically solving the Boltzmann equation with an overoccupied Glasma initial condition. 
Such a study was previously performed\footnote{There are studies of the Boltzmann equation with an expanding geometry but they have not included yet the $(1+f)$ Bose enhancement factor that is essential to match the classical Yang-Mills simulations to kinetic theory~\cite{Ruggieri:2015tsa,Greif:2016jeb}.} by Kurkela and Zhu~\cite{Kurkela:2015qoa}, who implemented the effective kinetic theory of Arnold, Moore and Yaffe~\cite{Arnold:2002zm} including both elastic  \ttt scattering and soft splitting processes. They however restricted their simulations to include only gluons. Further, their focus was to follow the evolution of the Glasma all the way to thermalization {\it a la} the bottom-up scenario. 

Our interest here is primarily the classical stage of the Glasma and we will perform simulations for a wider range of couplings than those presented in \cite{Kurkela:2015qoa}. Firstly, by including only elastic  \ttt processes, we will investigate how well these kinetic studies reproduce the self-similar scaling behavior of the gluon distribution, that has been obtained by the classical-statistical simulations.  We will employ the small angle approximation to the Boltzmann equation whereby, as shown by Landau~\cite{landau1936kinetic,lifshitz1981physical}, the collision integral is approximated by a Fokker-Planck like diffusion term\footnote{The Landau kinetic equation was first applied to the Glasma by Mueller \cite{Mueller:1999pi} and simulated numerically in~\cite{Bjoraker:2000cf}. However in Refs.~\cite{Mueller:1999pi,Bjoraker:2000cf}, the Bose enhancement factor, important for the description of overoccupied gluons, was not taken into account.}. We will demonstrate that a flow of gluon number to the infrared (IR) generates an IR enhancement that often interpreted as a transient Bose-Einstein condensate (BEC) in the literature. We will present results for both the longitudinal {\it and} transverse momentum distributions as well as the behavior of $P_L/P_T$, demonstrating the influence of this IR enhancement. We should emphasize at the outset that the IR enhancement is unlikely to capture key features of BEC formation as studied and quantified in detail for identical overoccupied conditions in scalar theories off-equilibrium~\cite{Berges:2014bba,Berges:2015ixa}.  

The generation of such an IR enhancement is unsurprising for a system of overoccupied gluons, as argued in a number of papers~\cite{Blaizot:2011xf,Blaizot:2013lga,Xu:2014ega}. What is surprising is that, when the BEC becomes manifest, our results reproduce important qualitative features of the classical-statistical simulations of the Glasma. If inelastic scattering processes played a large role at these early times, a BEC would never form and would not influence the dynamics of the Glasma. Our results therefore suggest that, at least for the range of weak couplings studied, 
inelastic processes may not be sufficiently strong at early times to hinder the formation of a transient BEC. 

We emphasize that our framework does not provide a first principles solution to some of the puzzling dynamics of the Glasma that we listed. It does not contain either $2\leftrightarrow 3$ scattering or the plasma instabilities that should be there in a purely kinetic picture. Further, it is unclear how one defines a BEC of gluons in QCD. Our result should instead be thought of as an effective description of the classical-statistical numerical simulations for the range of couplings in those studies. Both inelastic scattering processes and late time plasma instabilities, whose dynamics is in principle contained in the classical-statistical simulations, appear to be suppressed in the early-time evolution of the overoccupied Glasma. Our results therefore argue for a deeper understanding and characterization of the IR dynamics in the Glasma.  Key features of this dynamics are captured by the spatial string tension, which has a characteristic time dependence~\cite{Mace:2016svc} that is universal~\cite{Berges:2017igc}. It has been observed that the spatial string tension sets the scale for off-equilibrium sphaleron transitions~\cite{Mace:2016svc}; this latter quantity is essential for dynamical simulations~\cite{Mueller:2016ven,Mace:2016shq} of the Chiral Magnetic effect.

The other novel feature of our study is the kinetic description of quark distributions in the expanding Glasma. The Landau kinetic equations were studied for an isotropic distribution of overoccupied gluons along with quarks and anti-quarks \cite{Blaizot:2014jna}; our work extends this study to the expanding Glasma. We will show that quark distributions in the Glasma also obey self-similar scaling distributions with the same dynamical scaling exponents as those satisfied by the gluons. In an  accompanying paper, we applied the insights from this study to investigate photon production in the Glasma relative to the quark-gluon plasma~\cite{Berges:2017eom}. For completeness, we will study chemical equilibration in the Glasma--while interesting, the results here for the relative population of quarks and gluons will be modified by the collinear gluon splitting processes that also influence kinetic equilibrium. Their additional contribution will be explored in subsequent work. 

The paper is organized as follows. In the next section, we will outline the kinetic equations for the coupled system of gluons, quarks and anti-quarks in the Glasma. Our discussion here is a straightforward extension of that in \cite{Blaizot:2014jna}. Numerical results are discussed in section 3 for the Glasma with no quarks or anti-quarks  ($N_f=0$). We consider the time evolution of the longitudinal and transverse hard scales, the ratio of the longitudinal and transverse pressure, and the self-similarity of the gluon distribution. For very weak couplings, qualitative agreement is found with the classical-statistical simulations, while significant deviations are observed for larger couplings. In section 4, we extend our study to finite $N_f$. The time evolution of the quark distributions is studied for a range of initial occupancies. As noted, the self-similar behavior that characterizes the gluon distribution holds for quark distributions as well. The approach to chemical equilibration is studied and is shown to be sensitive to the initial quark occupancy. The final section contains a brief summary of our results and a discussion of future work. The numerical procedure is outlined in Appendix A and the features of the BEC formed are discussed in Appendix B.

\section{Kinetic equation for \ttt elastic scatterings in the Glasma}
\subsection{Small angle approximation for the Boltzmann equation}
We begin by defining the gluon distribution function as a number density in phase space averaged over all polarization and color states: 
\begin{equation}
f(\tau ,\bp ) = \frac{(2\pi)^3}{2(N_c^2-1)} \frac{dN_\text{gluon}}{d^3 x d^3 p} \, .
\end{equation} 
Similarly, for $N_f$-flavor massless and unpolarized quarks, the distribution function is defined as
\begin{equation}
F(\tau ,\bp ) = \frac{(2\pi)^3}{2N_c N_f} \frac{dN_\text{quark}}{d^3 x d^3 p} \, .
\end{equation}
We will assume that the system is boost-invariant in the longitudinal direction and uniform in the transverse direction, so that
$f$ and $F$ are functions of $\tau =\sqrt{t^2-z^2}$ and $\bp$ alone. 
Furthermore, we assume that anti-quarks have the same distribution as quarks. 
Each of these distributions satisfies a Boltzmann equation, defined as 
\begin{align}
\left( \frac{\partial}{\partial \tau} -\frac{p_z}{\tau} \frac{\partial}{\partial p_z} \right) f(\tau ,\bp ) 
 &= C_\text{gluon}[f,F] \, , \label{Boltzmann_gluon} \\
\left( \frac{\partial}{\partial \tau} -\frac{p_z}{\tau} \frac{\partial}{\partial p_z} \right) F(\tau ,\bp ) 
 &= C_\text{quark}[F,f] \, , \label{Boltzmann_quark}
\end{align}
where $-\frac{p_z}{\tau} \frac{\partial f}{\partial p_z}$ is the drift term describing the effect of the longitudinal expansion. 
For the collision terms $C_\text{gluon}[f,F]$ and $C_\text{quark}[F,f]$, we consider only \ttt elastic scatterings. Since gluons, quarks and anti-quarks scatter off each other, 
the collision terms of each of the equations depend on the distributions of the other,  which makes Eqs.~\eqref{Boltzmann_gluon} and \eqref{Boltzmann_quark} a coupled set of equations.
In the small angle approximation, the collision terms can be expressed as the sum of a diffusion term and a source term \cite{Blaizot:2014jna},
\begin{align}
C_\text{gluon}[f,F] &= -\nabla_\bp \cdot \bJ_\text{g} +S_\text{g} \, , \label{col_g0} \\
C_\text{quark}[F,f] &= -\nabla_\bp \cdot \bJ_\text{q} +S_\text{q} \, . \label{col_q0}  
\end{align}
Here $\bJ_\text{g}$ and $\bJ_\text{q}$ denote particle currents in momentum space and are given by
\begin{align}
\bJ_\text{g} &= -\frac{g^4}{4\pi} N_c \mathcal{L} \left[ I_a \nabla_\bp f +I_b \frac{\bp}{p} f \left(1+f \right) \right] \, , \label{Jg} \\
\bJ_\text{q} &= -\frac{g^4}{4\pi} C_F \mathcal{L} \left[ I_a \nabla_\bp F +I_b \frac{\bp}{p} F \left(1-F \right)\right] \, .
\end{align}
The source terms $S_\text{g}$ and $S_\text{q}$ are respectively
\begin{align}
S_\text{g} &= \frac{g^4}{4\pi} C_F N_f \mathcal{L} I_c \frac{1}{p} \left[ F(1+f)-f(1-F)\right] \, , \\
S_\text{q} &= -\frac{g^4}{4\pi} C_F^2 \mathcal{L} I_c \frac{1}{p} \left[ F (1+f)-f(1-F)\right] \, ,
\end{align}
where $C_F=(N_c^2-1)/(2N_c)$ is the quadratic Casimir in the fundamental representation of SU$(N_c)$. 
In obtaining these expressions, it has been assumed that $f(\tau ,\bp )=f(\tau ,-\bp )$ and $F(\tau ,\bp )=F(\tau ,-\bp )$.   
We have further defined the integrals 
\begin{equation}
I_a (\tau ) = \int\! \frac{d^3p}{(2\pi)^3} \left[ N_c f (1+f) +N_f F (1-F) \right] \, ,
\end{equation}
\begin{equation}
I_b (\tau ) = 2\int\! \frac{d^3p}{(2\pi)^3} \frac{1}{p}\left[ N_c f +N_f F  \right] \, ,
\end{equation}
and
\begin{equation}
I_c (\tau ) = \int\! \frac{d^3p}{(2\pi)^3} \frac{1}{p}\left[ f +F \right] \, .
\end{equation}
The integrand of $I_a$ is proportional to the density of possible scatterers that are enhanced or suppressed by the Bose or the Pauli factor respectively. 
The integral $I_b$ is related to the Debye mass scale $m_D$ as $m_D^2=2g^2 I_b$. 
The Coulomb logarithm $\mathcal{L}$ is a divergent integral that is regularized by cutoffs as
\begin{equation} \label{cl}
\mathcal{L} = \int_{q_\text{min}}^{q_\text{max}} \frac{dq}{q} = \log \frac{q_\text{max}}{q_\text{min}} \, .
\end{equation}
This IR divergence originates from the long range nature of the interaction and is regularized by the medium mass. 
We will  use the Debye mass scale
\begin{equation} \label{Debye}
m_D = \sqrt{2g^2 I_b (\tau )} 
\end{equation}
as the IR cutoff. Since $m_D$ is time dependent, so too is the Coulomb logarithm. 

The ultraviolet (UV) divergence, in contrast, is not physically meaningful since it arises from the small momentum expansion performed inside the momentum integrals that give the above form of the kinetic equations. The original collision integral does not possess this UV divergence because the distribution functions have finite support in momentum space. It is therefore natural to use the typical momentum scale of the distribution functions as the UV cutoff. 
Since the typical transverse momentum scale is larger than the longitudinal one in a longitudinally expanding Glasma, we will use the square root of the mean $\pperp^2$,
\begin{equation} \label{mean_pT}
\langle \pperp^2 \rangle
=\frac{\int \! d^3 p \, \pperp^2 f(\tau ,\pperp ,p_z )}{\int \! d^3 p \, f(\tau ,\pperp ,p_z )} \,,
\end{equation}
as the UV cutoff for the Coulomb logarithm.
In the overoccupied weak coupling regime of interest in this study, this scale is nearly constant and close to the saturation scale $Q_s$. We will therefore simply choose the UV cutoff to be $Q_s$, except in Sec.~\ref{subsec:equilibrium}.

\subsection{Rescaling for an overoccupied Glasma}
In the following, we will exploit the fact that the Glasma is highly overoccupied at early times. 
To simplify further analysis, we will rescale the gluon distribution as
\begin{equation} \label{def_ftilde}
\tilde{f} (\tau, \bp ) = \frac{f(\tau, \bp )}{f_0} \, ,
\end{equation}
with $f_0$ being the maximum value of the initial distribution $f(\tau_0,\bp )$.  We further introduce the rescaled time $\tilde{\tau}$,
\begin{equation}
\tilde{\tau} = \frac{1}{4\pi} (N_c g^2 f_0 )^2 \mathcal{L}_0 \, \tau \, ,
\end{equation}
with $\mathcal{L}_0 $ denoting the initial value of the Coulomb logarithm, $\mathcal{L}_0 =\mathcal{L} (\tau_0)$, determined by the initial distributions.
Since the quark occupancy is always smaller than one, we will not rescale it. 

In terms of the rescaled quantities, the kinetic equations can be expressed as 
\begin{equation} \label{kine_g1}
\left( \frac{\partial}{\partial \tilde{\tau}} -\frac{p_z}{\tilde{\tau}} \frac{\partial}{\partial p_z} \right) 
\tilde{f} ( \tilde{\tau} ,\bp )
= \tilde{\mathcal{L}} \nabla_\bp \cdot \left[ \tilde{I}_a \nabla_\bp \tilde{f}(\tilde{\tau} ,\bp ) 
+\frac{\bp}{p} \tilde{I}_b \tilde{f}(\tilde{\tau} ,\bp ) \left(f_0^{-1}+\tilde{f}(\tilde{\tau} ,\bp ) \right) \right] 
+\frac{C_F N_f}{N_c^2 f_0} \tilde{S} \, ,
\end{equation}
for gluons, and
\begin{equation} \label{kine_q1}
\left( \frac{\partial}{\partial \tilde{\tau}} -\frac{p_z}{\tilde{\tau}} \frac{\partial}{\partial p_z} \right) 
F ( \tilde{\tau} ,\bp )
= \frac{C_F}{N_c} \tilde{\mathcal{L}} \nabla_\bp \cdot \left[ \tilde{I}_a \nabla_\bp F(\tilde{\tau} ,\bp ) 
+\frac{\bp}{p} \tilde{I}_b F(\tilde{\tau} ,\bp ) \left(1-F(\tilde{\tau} ,\bp ) \right) \right] 
-\frac{C_F^2}{N_c^2} \tilde{S} \, ,
\end{equation}
for quarks, with the rescaled source term
\begin{equation}
\tilde{S} = \tilde{\mathcal{L}} \tilde{I}_c \frac{1}{p} \left[ F(f_0^{-1}+\tilde{f})-\tilde{f}(1-F)\right] \, .
\end{equation}
The corresponding rescaled integrals are 
\begin{equation}
\tilde{I}_a (\tilde{\tau} ) = \int\! \frac{d^3p}{(2\pi)^3} \left[ \tilde{f} (f_0^{-1}+\tilde{f}) +\frac{N_f}{N_c f_0^2} F (1-F) \right] \, ,
\end{equation}
\begin{equation}
\tilde{I}_b (\tilde{\tau} ) = 2\int\! \frac{d^3p}{(2\pi)^3} \frac{1}{p}\left[ \tilde{f} +\frac{N_f}{N_c f_0} F  \right] \, ,
\end{equation}
\begin{equation}
\tilde{I}_c (\tilde{\tau} ) = \int\! \frac{d^3p}{(2\pi)^3} \frac{1}{p}\left[ \tilde{f} +\frac{1}{f_0} F \right] \, ,
\end{equation}
and we have introduced a normalized Coulomb logarithm defined as 
\begin{equation} \label{cl_norm}
\tilde{\mathcal{L}} (\tilde{\tau}) = \frac{\mathcal{L} (\tau)}{\mathcal{L}_0 } \, .
\end{equation}
In terms of the rescaled integral $\tilde{I}_b$, the Debye mass is expressed as
\begin{equation}
m_D^2 = 2N_c g^2 f_0 \tilde{I}_b (\tilde{\tau} ) \, .
\end{equation}

Hereafter, we shall fix $N_c=N_f=3$. 
The relevant physical parameters for this system are the coupling $g$, the initial occupancy $f_0$, 
and the initial time $\tilde{\tau}_0$. To ensure that a kinetic approach based on a perturbative collision kernel is valid, both $g$ and  $g^2 f_0$ must be sufficiently small. 

Comparing, Eqs.~\eqref{kine_g1} and \eqref{kine_q1}, we can make the following observations. 
\begin{itemize}
\item Since $C_F/N_c=4/9 <1$, the effect of diffusion is weaker for quarks than gluons.    
\item If the initial state is highly occupied by gluons, \textit{i.e.} $f_0\gg 1$, the source term is much smaller than the diffusion term for gluons, while for quarks the source term is always as important as the diffusion term at least parametrically.  
\end{itemize}

\subsection{Conservation laws}
One can easily confirm that the collision terms, as approximated by Eqs.~\eqref{col_g0} and \eqref{col_q0}, conserve both the total particle number density
\begin{equation} \label{ntot}
n(\tau ) = n_\text{g} (\tau ) +n_\text{q} (\tau ) \, ,
\end{equation} 
and the total energy density
\begin{equation} \label{etot}
\mathcal{E}(\tau ) = \mathcal{E}_\text{g} (\tau ) +\mathcal{E}_\text{q} (\tau ) \, ,
\end{equation} 
where the particle number density and the energy density, respectively, of each species are
\begin{equation} \label{n_g}
n_\text{g}(\tau ) = 2(N_c^2-1) \int \! \frac{d^3p}{(2\pi)^3} f(\tau ,\bp ) \, ,
\end{equation} 
\begin{equation}
n_\text{q}(\tau ) = 4N_c N_f \int \! \frac{d^3p}{(2\pi)^3} F(\tau ,\bp ) \, ,
\end{equation} 
\begin{equation} 
\mathcal{E}_\text{g}(\tau ) = 2(N_c^2-1) \int \! \frac{d^3p}{(2\pi)^3} \, p f(\tau ,\bp ) \, ,
\end{equation} 
and
\begin{equation} 
\mathcal{E}_\text{q}(\tau ) = 4N_c N_f \int \! \frac{d^3p}{(2\pi)^3} \, p F(\tau ,\bp ) \, . 
\end{equation} 
In the presence of the expansion term, the conservation laws are modified to
\begin{equation} \label{n_conv}
\frac{d}{d\tau} (\tau n) = 0 \, ,
\end{equation}
and 
\begin{equation} \label{ene_conv}
\frac{d\mathcal{E}}{d\tau} = -\frac{\mathcal{E} +\mathcal{P}_L}{\tau } \, ,
\end{equation}
where $\mathcal{P}_L$ is the longitudinal  pressure given by the sum of
\begin{equation}
\mathcal{P}_{L,g} (\tau ) = 2(N_c^2-1) \int \! \frac{d^3p}{(2\pi)^3} \frac{p_z^2}{p} f(\tau ,\bp ) \, ,
\end{equation}
and
\begin{equation}
\mathcal{P}_{L,q} (\tau ) = 4N_c N_f \int \! \frac{d^3p}{(2\pi)^3} \frac{p_z^2}{p} F(\tau ,\bp ) \, .
\end{equation}
For later convenience, we also introduce the transverse pressure 
\begin{equation}
\mathcal{P}_{T,g} (\tau ) = 2(N_c^2-1) \int \! \frac{d^3p}{(2\pi)^3} \frac{p_x^2 +p_y^2}{2p} f(\tau ,\bp ) \, , \hspace{10pt}
\mathcal{P}_{T,q} (\tau ) = 4N_c N_f \int \! \frac{d^3p}{(2\pi)^3} \frac{p_x^2 +p_y^2}{2p} F(\tau ,\bp ) \, .
\end{equation}

\subsection{Initial conditions}
For a given initial distribution $f(\tau_0 ,\bp )$, the initial time $\tau_0$ which is consistent with that initial distribution can be specified by the following argument. 
The number density of hard gluons produced immediately after a heavy-ion collision can be expressed as~\cite{Mueller:1999fp}
\begin{equation}
n_\text{hard} (\tau ) = c\, \frac{(N_c^2-1) Q_s^3}{4\pi^2 N_c \alpha_s}\,\frac{1}{Q_s\tau} \, ,
\label{nhard}
\end{equation}
where $c$ is the gluon liberation coefficient, which can be estimated by using classical Yang-Mills simulations of the Glasma \cite{Krasnitz:2000gz,Krasnitz:2003jw,Lappi:2003bi}. We employ the value $c=1.1$ given in \cite{Lappi:2007ku}.  
As long as elastic scattering dominates amongst hard gluons, the number density decreases in time as $1/\tau$ due to the longitudinal expansion. 
Since the gluon number density is dominated by hard gluons at early times, the expression \eqref{nhard} should agree with $n_g$ given by \eqref{n_g}. 
By this condition, we can find the initial time $\tau_0$ that is consistent with the initial distribution $f_0 (\bp )=f(\tau_0 ,\bp )$ as 
\begin{align}
Q_s \tau_0 
&= \frac{c}{2\pi N_c g^2} \frac{Q_s^3}{\int \! \frac{d^3 p}{(2\pi)^3} f_0 (\bp ) } \, .
\end{align}
The rescaled time that corresponds to this initial time is
\begin{align} 
Q_s \tilde{\tau}_0 
&= \frac{c}{8\pi^2} N_c \, g^2 f_0 \mathcal{L}_0 \frac{Q_s^3}{\int \! \frac{d^3 p}{(2\pi)^3} \tilde{f}_0 (\bp ) } \, .
\label{tildetau0}
\end{align}

\section{Numerical results I. $N_f=0$} \label{sec:gluons_results}
In this section, we will show numerical results for a pure gluon system ($N_f=0$). 
We will employ a family of initial gluon distributions, 
\begin{equation} \label{ini_dist}
f( \tau_0 ,\pperp ,p_z ) = \frac{n_0}{g^2} e^{-\left[ \pperp^2 +(\xi_0 p_z )^2 \right]/Q_s^2} \, ,
\end{equation}
where $\pperp =\sqrt{p_x^2+p_y^2}$ is the transverse momentum, the parameter $\xi_0$ characterizes the anisotropy of the initial distribution and $n_0 =g^2 f_0 $ specifies the overoccupancy.  
For the kinetic description to be valid, $n_0$ must be sufficiently small, as it controls the effective interaction strength.   

Substituting this initial distribution into Eq.~\eqref{tildetau0}, one obtains the initial time
\begin{equation}
Q_s \tilde{\tau}_0 
= \frac{3c}{\sqrt{\pi}} \mathcal{L}_0 \, n_0\, \xi_0 \, ,
\end{equation}
where the initial value of the Coulomb logarithm is
\begin{equation}
\mathcal{L}_0 = -\frac{1}{2} \log \left( \frac{3n_0}{\pi^2} \frac{1}{\sqrt{\xi_0^2-1}} \arctan \sqrt{\xi_0^2-1} \right) \, .
\end{equation}
In the following, we will mostly use $n_0=0.1$. The corresponding numerical values of the initial time are $Q_s \tilde{\tau}_0 \simeq 0.33$, 0.74 and 1.7 for $\xi_0=1$, 2 and 4, respectively. 
For simplicity, we shall typically use the initial time of  $Q_s \tilde{\tau}_0 =1$ except in Sec.~\ref{subsec:equilibrium}.

To solve the kinetic equation in Eq.~\eqref{kine_g1} numerically, we use $p=\sqrt{\pperp^2 +p_z^2}$ and $\kappa =\cos \theta =p_z/p$ as the momentum variables. 
For $p$, we discretize the range $p_\text{min} \leq p \leq p_\text{max}$ nonuniformly into $N_p$ grid points. 
For the angular variable $\kappa$, 
it is sufficient to consider the range $0\leq \kappa \leq 1$, which is divided into $N_\kappa$ grid points. 
Unless otherwise noted, we use $p_\text{min}/Q_s =10^{-2}$, $p_\text{max}/Q_s =8$, $N_p=500$ and $N_\kappa =256$. 
We have confirmed that our results are insensitive to changes of these parameters. 
The details of the numerical method are discussed in Appendix \ref{sec:num_method}.

\subsection{Self-similar evolution} \label{subsec:gluon_results}
We will begin with a brief review of the classical regime (of gluon occupancies greater than unity) in the bottom-up thermalization scenario ($Q_s^{-1} \ll \tau \ll \alpha_s^{-3/2} Q_s^{-1}$)  \cite{Baier:2000sb}.
In this time range, the number of gluons is dominated by hard gluons whose transverse momenta are the order of the saturation scale $Q_s$, and quarks and gluons scatter at small angles with typical momentum transfers $\sim m_D$. 
The initial density of hard gluons is parametrically $Q_s^3/\alpha_s$, and it decreases in time due to the longitudinal expansion as
\begin{equation} \label{para_Nh}
n_\text{hard} \sim \frac{Q_s^3}{\alpha_s Q_s \tau} \, .
\end{equation} 
The numerical coefficient for this relation is given by Eq.~\eqref{nhard}; however, it is not necessary for this qualitative discussion. 
The Debye mass can be estimated as
\begin{equation} \label{para_mD}
m_D^2 \sim \alpha_s \int \! d^3 p \, \frac{f_\text{hard}(p)}{p} \sim \alpha_s \frac{n_\text{hard}}{Q_s} \sim \frac{Q_s^2}{Q_s \tau} \, . 
\end{equation}
Since the exchanged momentum is much smaller than the saturation scale, the typical transverse momentum of hard gluons stays around $Q_s$. 
The small angle scatterings primarily modify the typical longitudinal momentum, which we call $p_z$, by providing random kicks of order $m_D$.
Since these random and incoherent scatterings can be regarded as a diffusion process, the typical longitudinal momentum obeys
\begin{equation} \label{para_diff}
p_z^2 \sim N_\text{col} m_D^2 \, .
\end{equation} 
Here, $N_\text{col}$ is the number of the collisions. Using $f_\text{hard} \sim n_\text{hard} /(Q_s^2 p_z)$, one can estimate it as 
\begin{equation} \label{para_Ncoll}
\frac{N_\text{col}}{\tau} \sim \frac{dN_\text{col}}{d\tau} \sim \sigma n_\text{hard} (1+f_\text{hard}) \sim \sigma \frac{n_\text{hard}^2}{Q_s^2 p_z} \, ,
\end{equation}
where $(1+f_\text{hard})$ is the Bose enhancement factor and $\sigma \sim \alpha_s^2 /m_D^2$ is the cross section. 
By combining \eqref{para_Nh}, \eqref{para_diff} and \eqref{para_Ncoll}, we obtain
\begin{equation}
p_z \sim Q_s (Q_s \tau)^{-1/3} \, ,
\end{equation} 
and
\begin{equation}
f_\text{hard} \sim \frac{n_\text{hard}}{Q_s^2 p_z} \sim \alpha_s^{-1} (Q_s \tau)^{-2/3} \, .
\end{equation}
These results can be summarized by the scaling relation
\begin{equation} \label{scaling}
f(\tau ,\pperp ,p_z ) = (Q_s \tau)^{-2/3} f_S \left( \pperp ,(Q_s \tau)^{1/3} p_z \right) \, ,
\end{equation}
where $f_S$ is a scaling function. 
As noted previously, this scaling law has been confirmed by  large scale classical-statistical real-time lattice simulations \cite{Berges:2013eia,Berges:2013fga}. 
The keys to the above derivation are the longitudinal expansion, the use the diffusion relation Eq.~\eqref{para_diff} and the Bose enhancement factor in Eq.~\eqref{para_Ncoll}. 
All of these features are included in the kinetic equation with the small angle approximation Eq.~\eqref{kine_g1}. 
Indeed, the diffusion constant $I_a$ can be interpreted as $dN_\text{col}/d\tau$. Therefore we anticipate that bottom-up thermalization in the overoccupied Glasma is well described by the kinetic equation \eqref{kine_g1}. 

In the above discussion, the soft splitting processes are not considered. Although those number-changing processes are naively higher order in the coupling, they give the leading order contributions because of the collinear enhancement \cite{Arnold:2002zm}. However, their main effect is producing soft gluons that have transverse momentum much smaller than $Q_s$, and the self-similar scaling behavior shown by the hard gluons is not affected by them. In the bottom-up thermalization scenario, the soft splitting processes play a crucial role in the later stage $\tau \gg \alpha_s^{-3/2} Q_s^{-1}$ \cite{Baier:2000sb}.

\begin{figure}[tb]
 \begin{center}
  \includegraphics[clip,width=8cm]{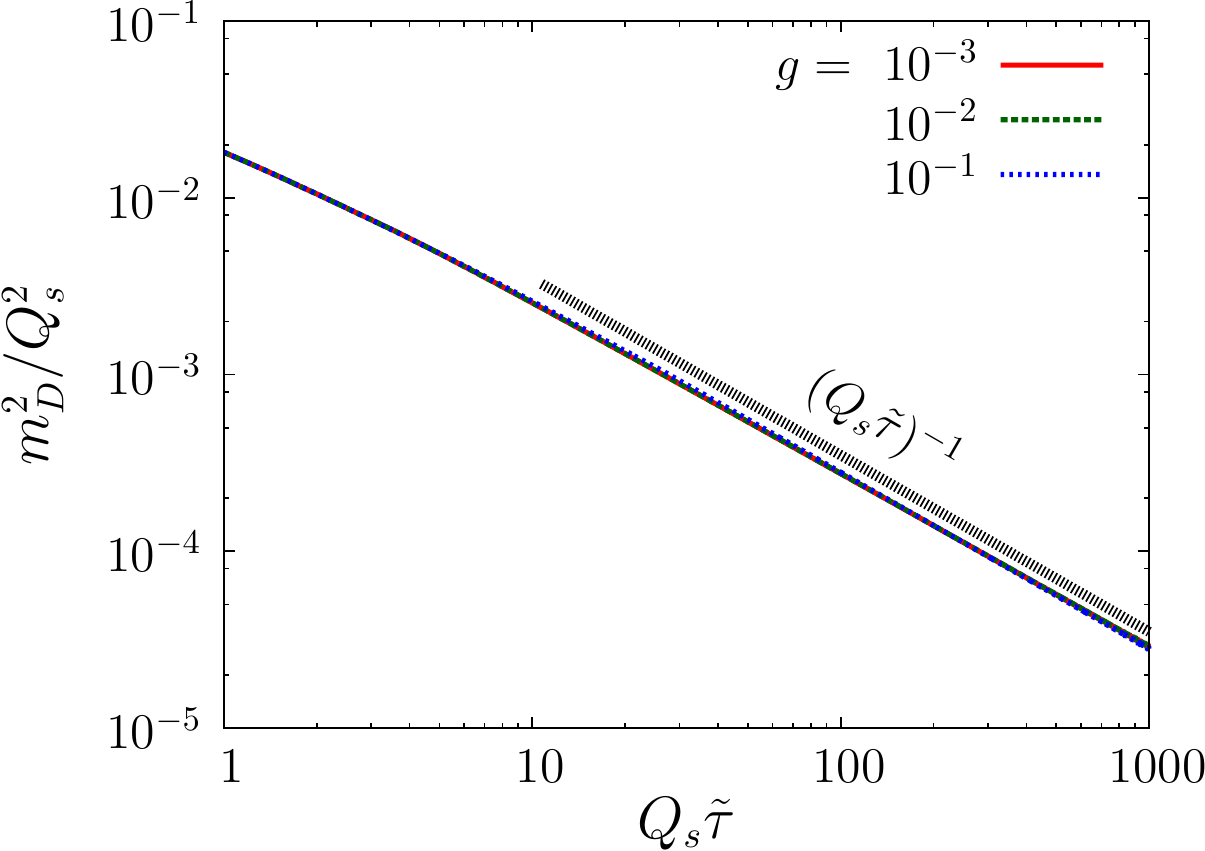} \vspace{-10pt}
  \caption{The time evolution of the Debye mass squared for different values of the coupling. The three curves lie on top of each other. 
 Except at early times, the Debye mass squared decreases as $1/\tau$.}
  \label{fig:Debye}
 \end{center}
\end{figure}

In Fig.~\ref{fig:Debye}, the Debye mass squared is plotted as a function of the rescaled time.
Results for several values of the coupling, $g=10^{-3}$, $10^{-2}$ and $10^{-1}$, are compared.  
Other parameters are $n_0 =0.1$, $\xi_0=2$ and the initial time\footnote{%
These parameters correspond to the initial time $Q_s \tau_0 \simeq 52$ in the original time variable.} $Q_s \tilde{\tau}_0=1$.
Since we fix $n_0$, the Debye mass is independent of the coupling in the classical scaling regime. 
The $1/\tau$ decrease in Eq.~\eqref{para_mD} can be confirmed clearly in the simulations. 

\begin{figure}[tb]
\begin{tabular}{cc}
 \begin{minipage}{0.5\hsize}
  \begin{center}
   \includegraphics[clip,width=8cm]{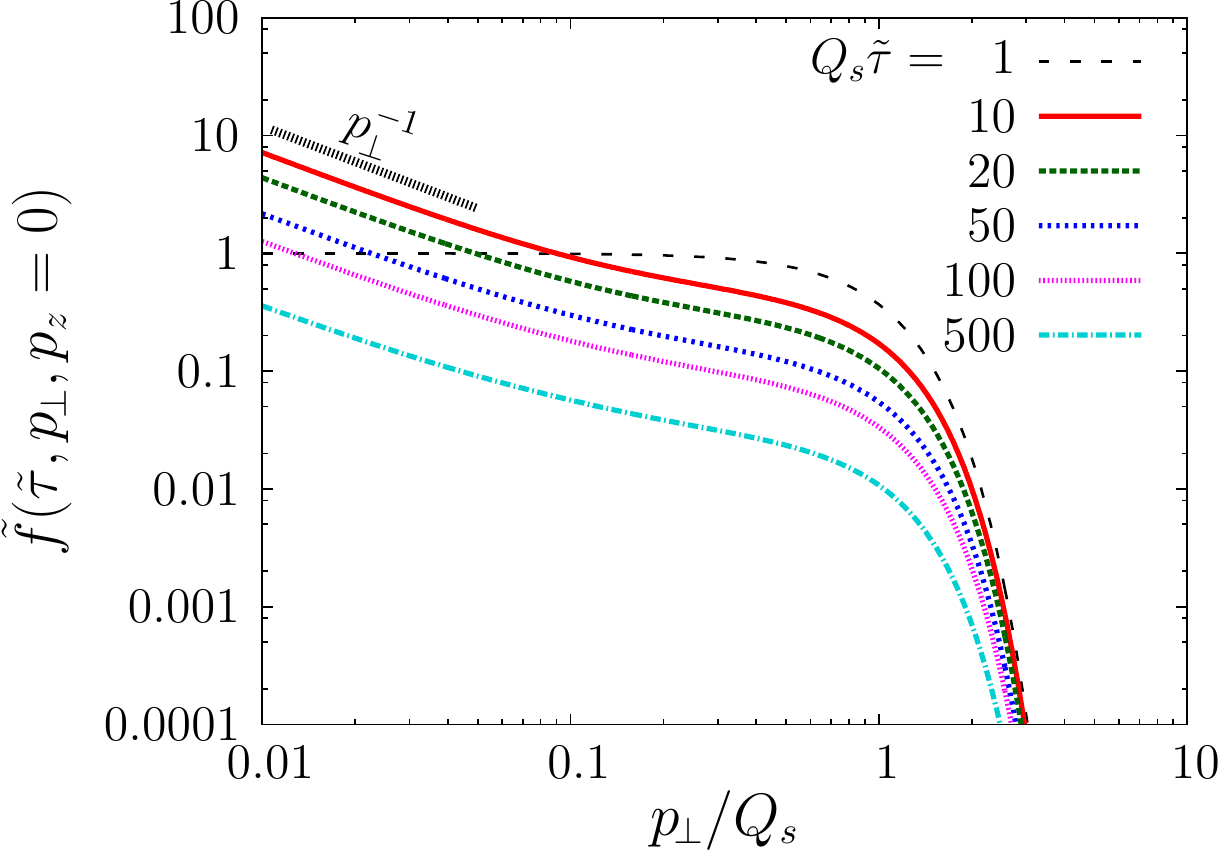} 
  \end{center}
 \end{minipage} &
 \begin{minipage}{0.5\hsize}
  \begin{center}
   \includegraphics[clip,width=8cm]{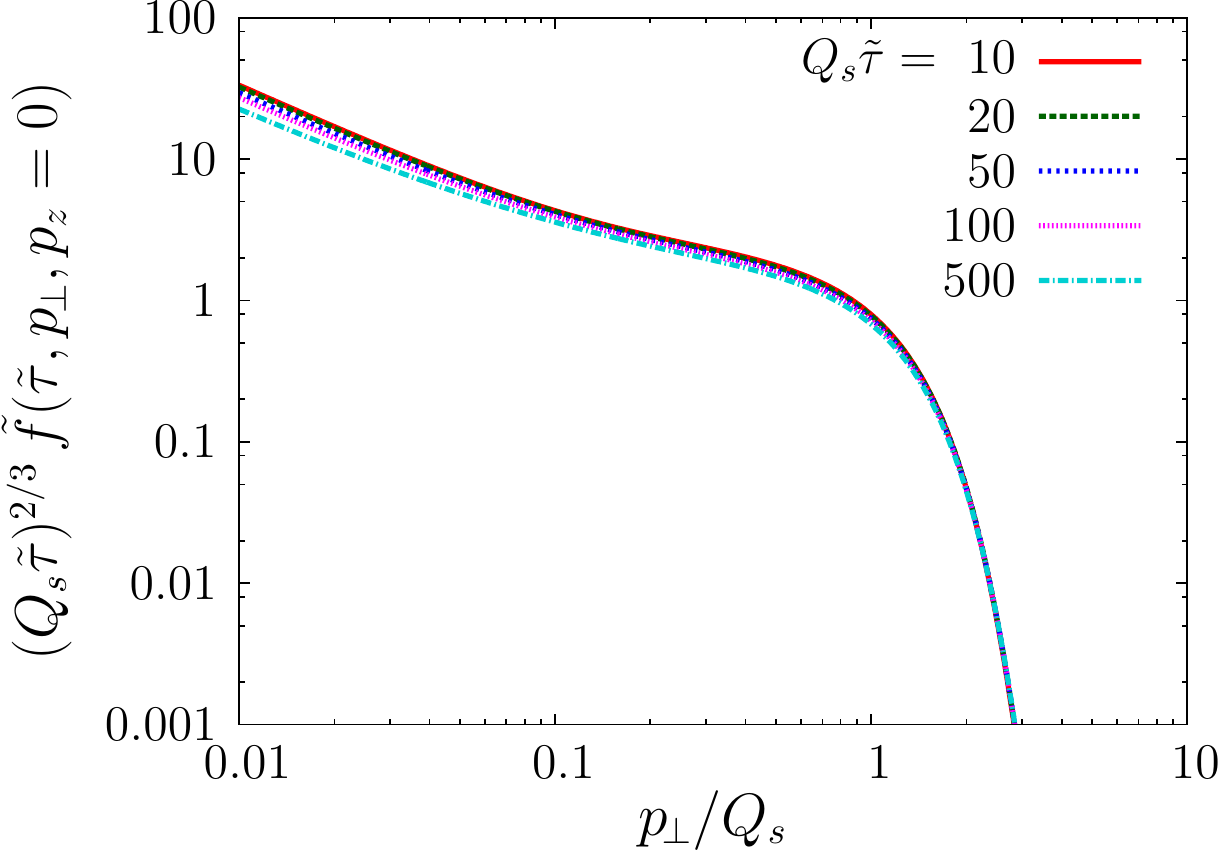} 
  \end{center}
 \end{minipage} 
\end{tabular}
\caption{The time evolution of the transverse momentum distribution evaluated at $p_z=0$.
Left panel: original distribution. Right panel: rescaled distribution.
For $\pperp \gtsim Q_s$, the scaling behavior shown in Eq.~\eqref{scaling} is well satisfied.}
\label{fig:distTg1}
\end{figure}

The time evolution of the rescaled distribution $\tilde{f}=f/f_0$ evaluated at $p_z=0$ is plotted as a function of the transverse momentum in the left panel of Fig.~\ref{fig:distTg1}. 
The parameters used for this computation are $g=10^{-3}$, $n_0=0.1$ ($f_0=10^5$), $\xi_0=2$ and $Q_s \tilde{\tau}_0=1$.
We note that in the infrared, the distribution shows $1/\pperp$ behavior. This result reproduces qualitatively\footnote{In detail, we note that the shape of the distribution near the hard scale is very similar to that observed for the scalar theory discussed at length in \cite{Berges:2015ixa}. In the classical-statistical simulations of the Glasma, this ``bump" was not seen for the times studied. We thank Juergen Berges for this observation.} those for $\pperp$ distributions from the classical-statistical simulations of the expanding Glasma~\cite{Berges:2013eia,Berges:2013fga}. In this kinetic framework, the emergence of the $1/\pperp$ behavior in the Glasma is a consequence of the onset of the transient Bose-Einstein condensate \cite{Blaizot:2013lga,Blaizot:2014jna}. The onset of the BEC within our approximation is discussed in Appendix \ref{sec:BEC}. For a proper description of the IR soft region, number-changing inelastic processes are essential \cite{Blaizot:2016iir}; further, it likely requires that one goes beyond a conventional kinetic framework and considers resummations along the lines discussed in~\cite{Orioli:2015dxa}. Our primary interest here lies in the universal scaling behavior that the hard modes show.

\begin{figure}[tb]
\begin{tabular}{cc}
 \begin{minipage}{0.5\hsize}
  \begin{center}
   \includegraphics[clip,width=8cm]{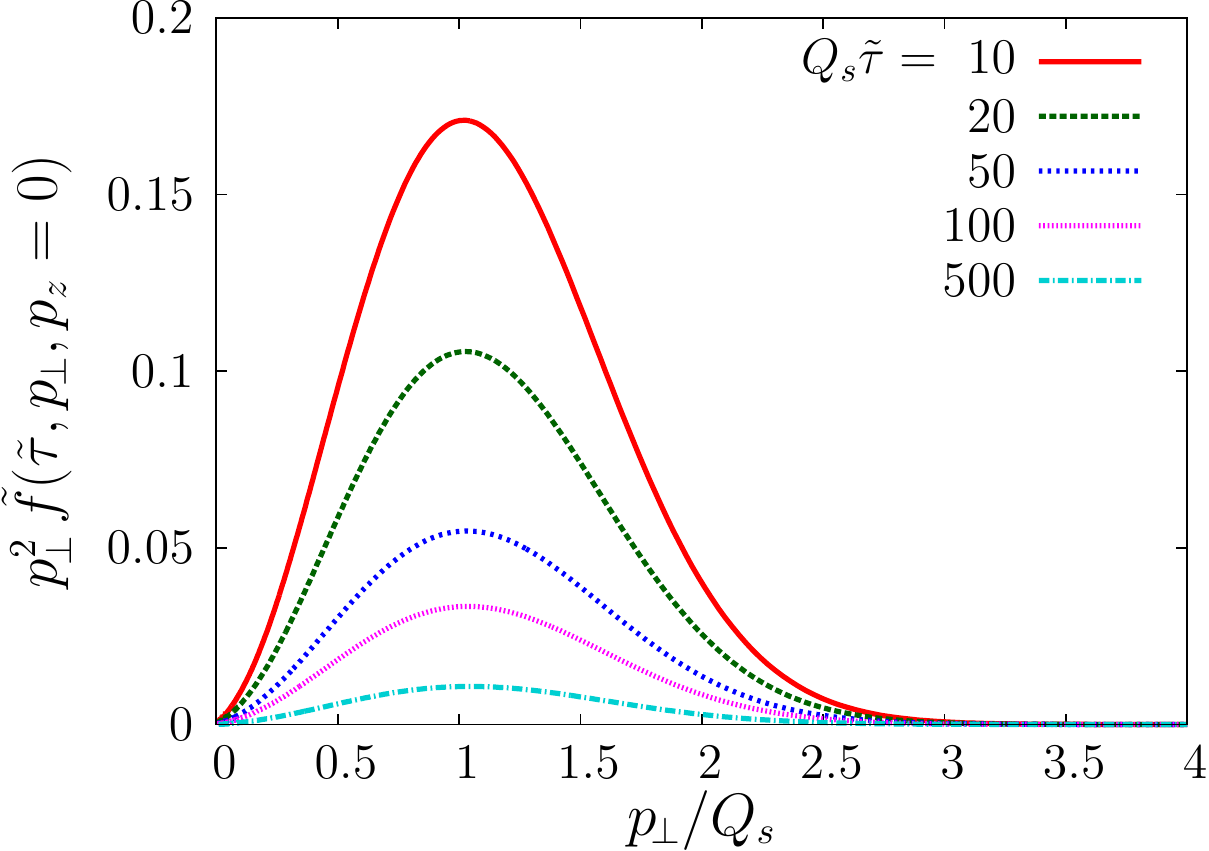} 
  \end{center}
 \end{minipage} &
 \begin{minipage}{0.5\hsize}
  \begin{center}
   \includegraphics[clip,width=8cm]{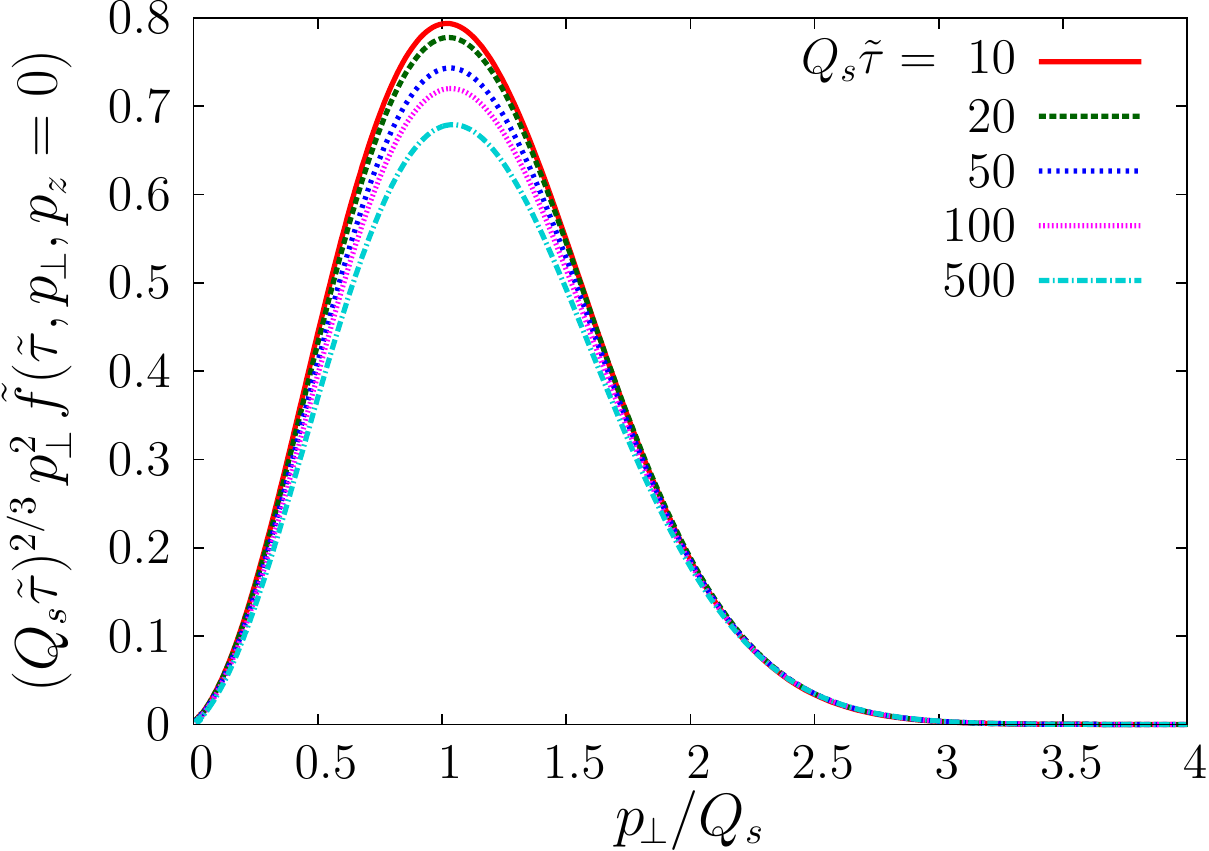} 
  \end{center}
 \end{minipage} 
\end{tabular}
\caption{The second moment of the transverse distribution, $\pperp^2 \tilde{f} (\tilde{\tau}, \pperp ,0)$, at different times.
Left panel: original distribution. Right panel: rescaled distribution.
The typical transverse momentum scale stays at $\sim Q_s$. }
\label{fig:distTg2}
\end{figure}

\begin{figure}[tb]
\begin{tabular}{cc}
 \begin{minipage}{0.5\hsize}
  \begin{center}
   \includegraphics[clip,width=8cm]{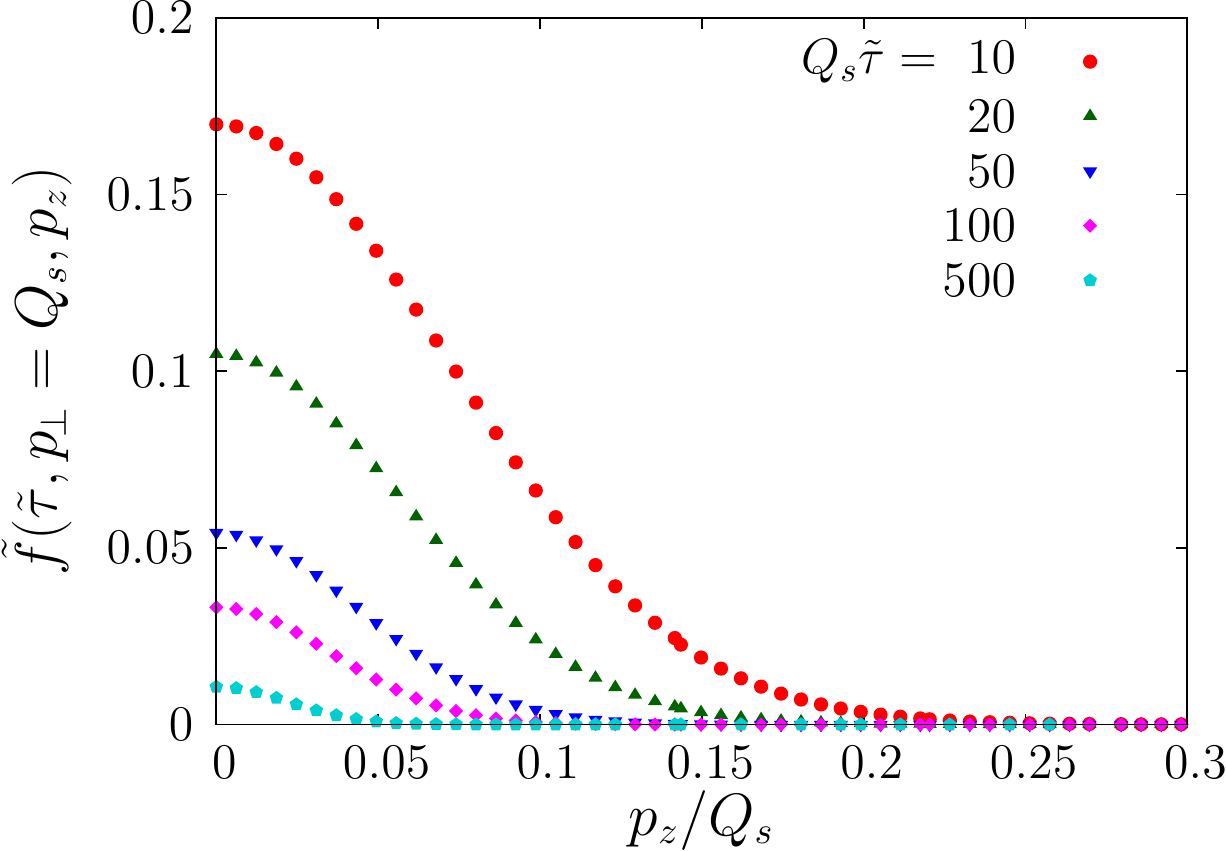}
  \end{center}
 \end{minipage} &
 \begin{minipage}{0.5\hsize}
  \begin{center}
   \includegraphics[clip,width=8cm]{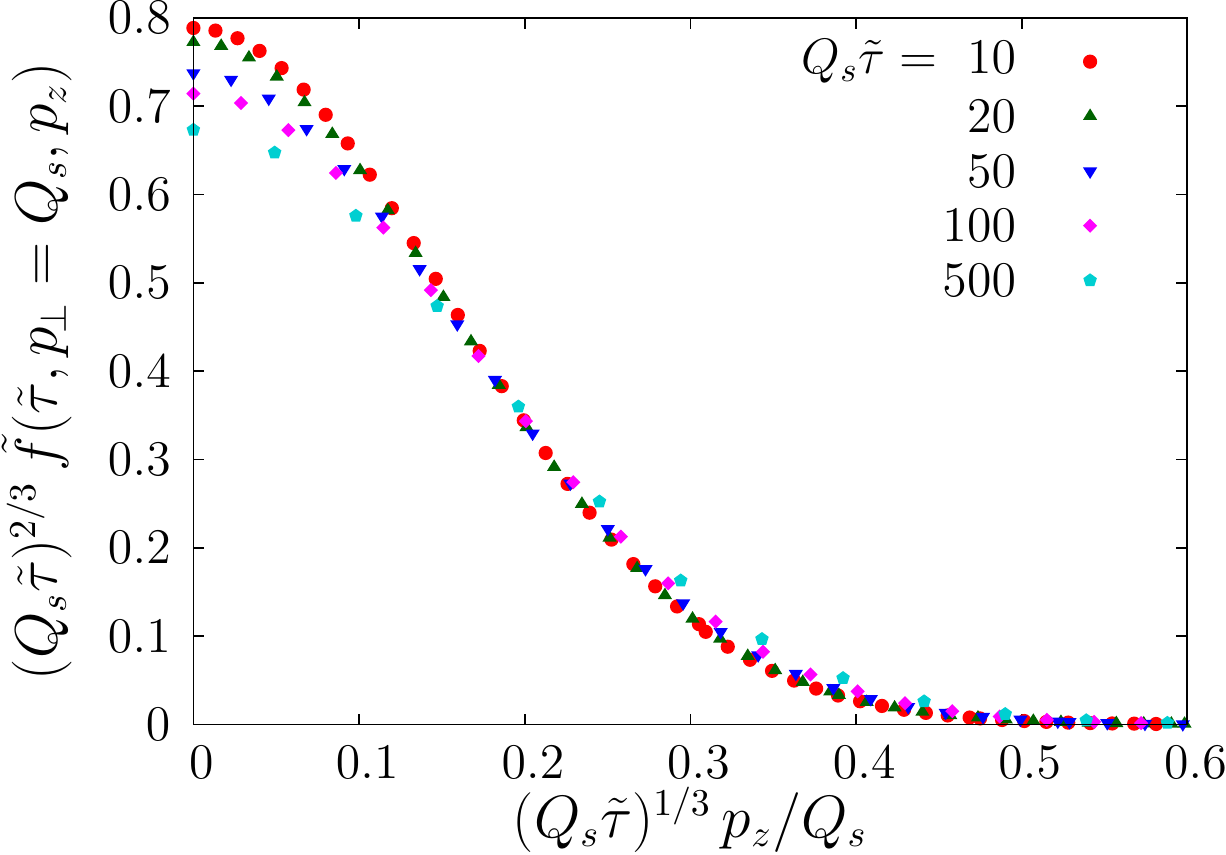}
  \end{center}
 \end{minipage} 
\end{tabular}
\caption{The time evolution of the longitudinal momentum distribution evaluated at $\pperp =Q_s$.
Left panel: original distribution. Right panel: rescaled distribution. }
\label{fig:distLg1}
\end{figure}

In the right panel of Fig.~\ref{fig:distTg1}, we plot the transverse momentum distribution multiplied by the scaling factor $(Q_s \tilde{\tau})^{2/3}$. 
For hard momenta $\pperp \gtsim Q_s$, the curves at different times  nicely overlap confirming the scaling behavior suggested by Eq.~\eqref{scaling}. 
Figure \ref{fig:distTg2} shows similar plots for the second moment of the transverse distribution, $\pperp^2 \tilde{f} (\tilde{\tau}, \pperp ,0)$. 
From these plots, one can confirm that the typical transverse momentum scale is of the order of $Q_s$. In Fig.~\ref{fig:distLg1}, the time evolution of the longitudinal momentum distribution evaluated at $\pperp =Q_s$ is plotted. 
The left panel shows the original distribution, while the right panel shows the distribution multiplied by the scaling factor $(Q_s \tilde{\tau})^{2/3}$ as a function of the rescaled momentum $(Q_s \tilde{\tau})^{1/3} p_z$.  Again, the rescaled distributions for vastly different times lie on top of each other. 
These results confirm that the nonequilibrium evolution of weakly coupled, overoccupied gluons in a longitudinally expanding system follows the bottom-up thermalization scenario. 

\begin{figure}[tb]
\begin{tabular}{cc}
 \begin{minipage}{0.5\hsize}
  \begin{center}
   \includegraphics[clip,width=7.4cm]{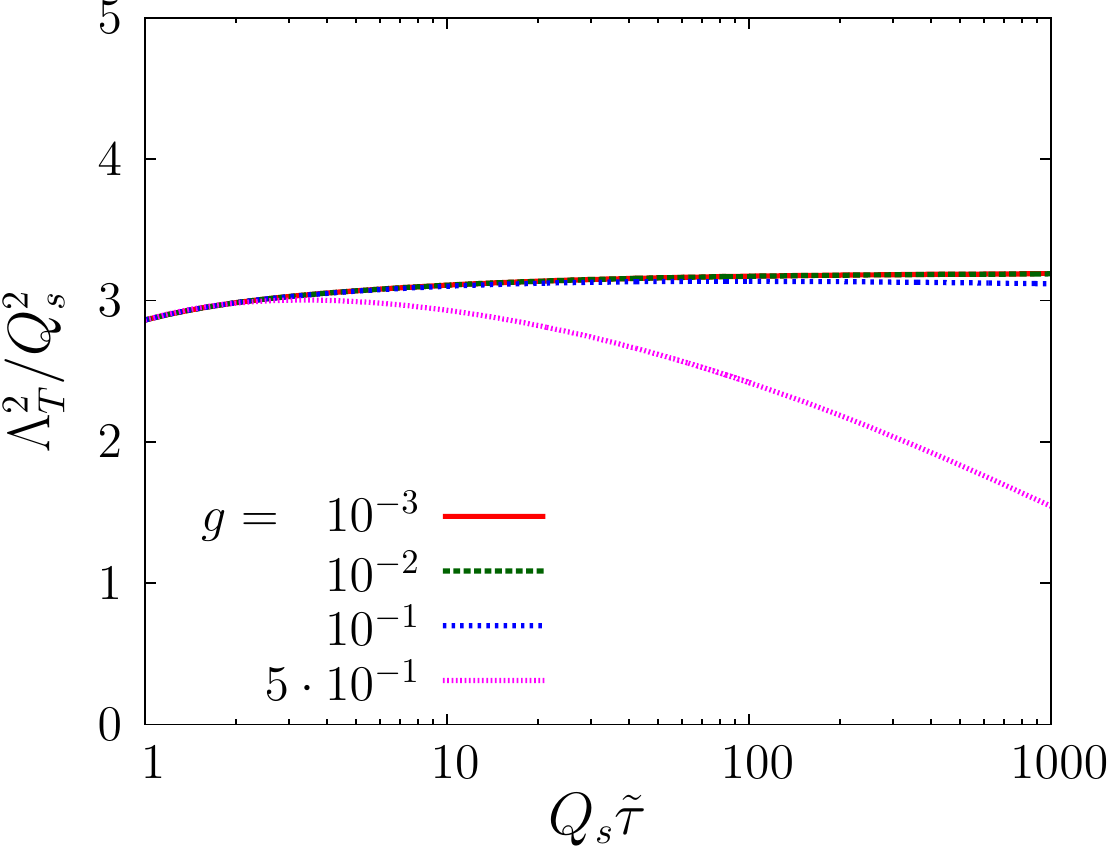}
  \end{center}
 \end{minipage} &
 \begin{minipage}{0.5\hsize}
  \begin{center}
   \includegraphics[clip,width=8.2cm]{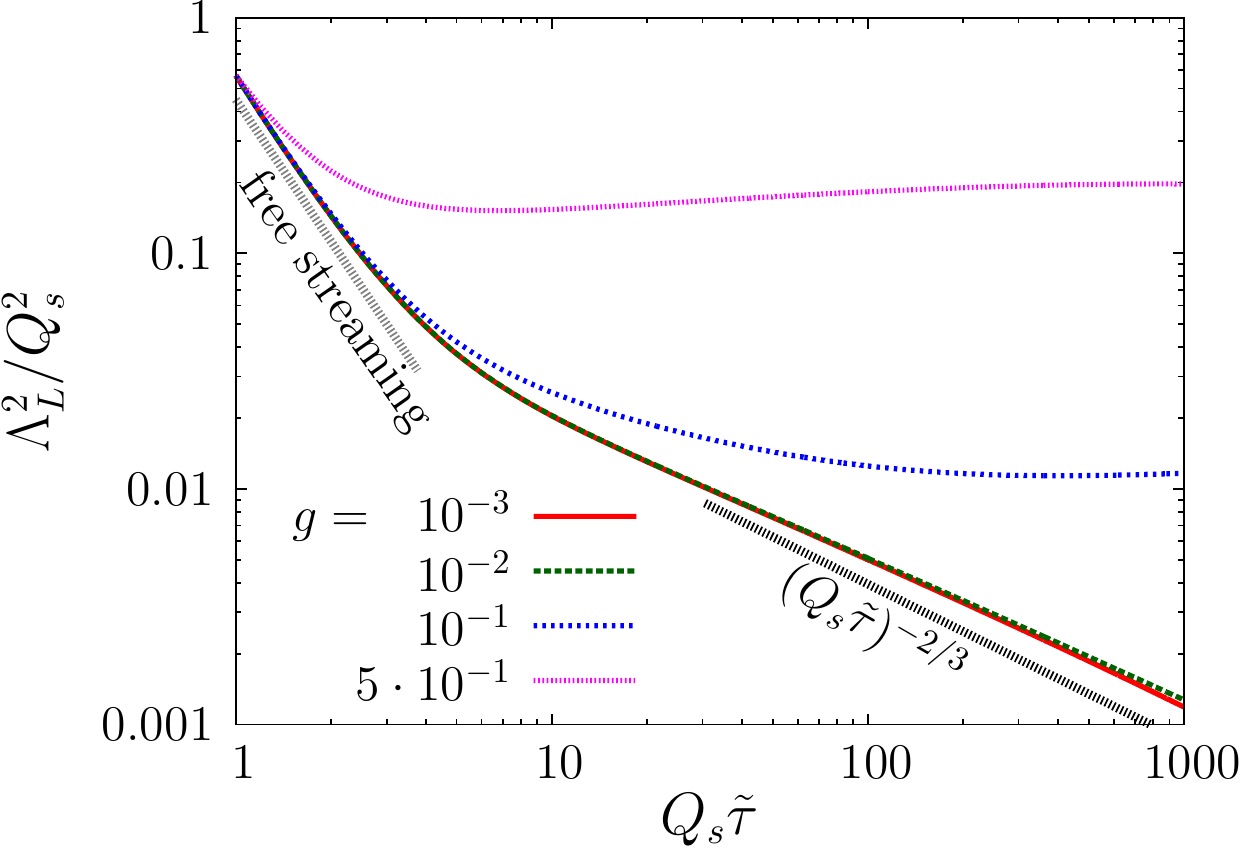}
  \end{center}
 \end{minipage} 
\end{tabular}
\caption{The time evolution of the transverse hard scale (left panel) and the longitudinal hard scale (right panel). Several values of the coupling are compared. 
   The occupancy measure is fixed to the value $n_0=0.1$. In the right figure, the free streaming behavior $\Lambda_L^2 (\tau) \sim (Q_s \tau )^{-2}$ and the classical scaling behavior $\sim (Q_s \tau )^{-2/3}$ are indicated by gray and black dashed lines respectively.}
  \label{fig:hs_g1}
\end{figure}

We also numerically compute the perturbative expressions for the hard scale observables \cite{Berges:2013fga}, 
\begin{equation} \label{hsT}
\Lambda_T^2 (\tau)
= \frac{\int\! d^3 p \, 2\pperp^2 |\bp| f(\tau, \bp)}{\int\! d^3 p \, |\bp| f(\tau, \bp)}
\end{equation}
and
\begin{equation} \label{hsL}
\Lambda_L^2 (\tau)
= \frac{\int\! d^3 p \, 4p_z^2 |\bp| f(\tau, \bp)}{\int\! d^3 p \, |\bp| f(\tau, \bp)} \, .
\end{equation}
In the scaling regime, the longitudinal hard scale is expected to behave as $\Lambda_L^2 (\tau) \sim (Q_s \tau )^{-2/3}$, while the transverse hard scale stays nearly constant. Their respective behavior is confirmed by the numerical results shown in Fig.~\ref{fig:hs_g1}, where the hard scales are plotted as a function of time for different values of the coupling. 
For the longitudinal hard scale, the free streaming behavior, $\Lambda_L^2 (\tau) \sim (Q_s \tau )^{-2}$, is seen at early times.
At later times, the system is driven to the classical scaling regime indicated by the temporal exponent $-2/3$ for sufficiently weak coupling, $g \ltsim 10^{-2}$. On the other hand,  results for the larger couplings (smaller occupancies) clearly deviate from the anticipated classical scaling behavior. 

\begin{figure}[tb]
 \begin{center}
  \includegraphics[clip,width=8cm]{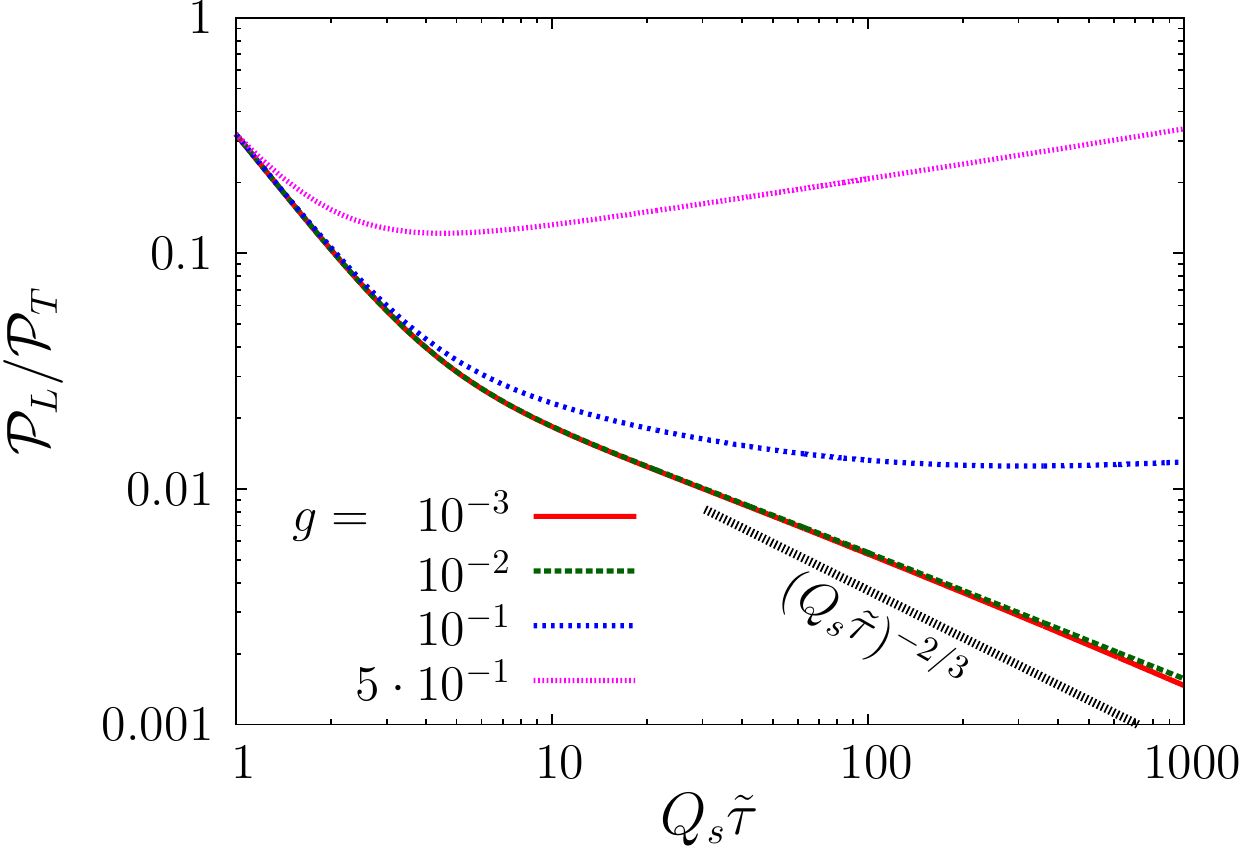} \vspace{-10pt}
  \caption{Time evolution of the bulk anisotropy for several values of the coupling constant. 
   A significant deviation from the temporal power law $(Q_s \tilde{\tau})^{-2/3}$ is observed even for weak couplings.}
  \label{fig:press_g1}
 \end{center}
\end{figure}

\begin{figure}[tb]
 \begin{center}
  \includegraphics[clip,width=8cm]{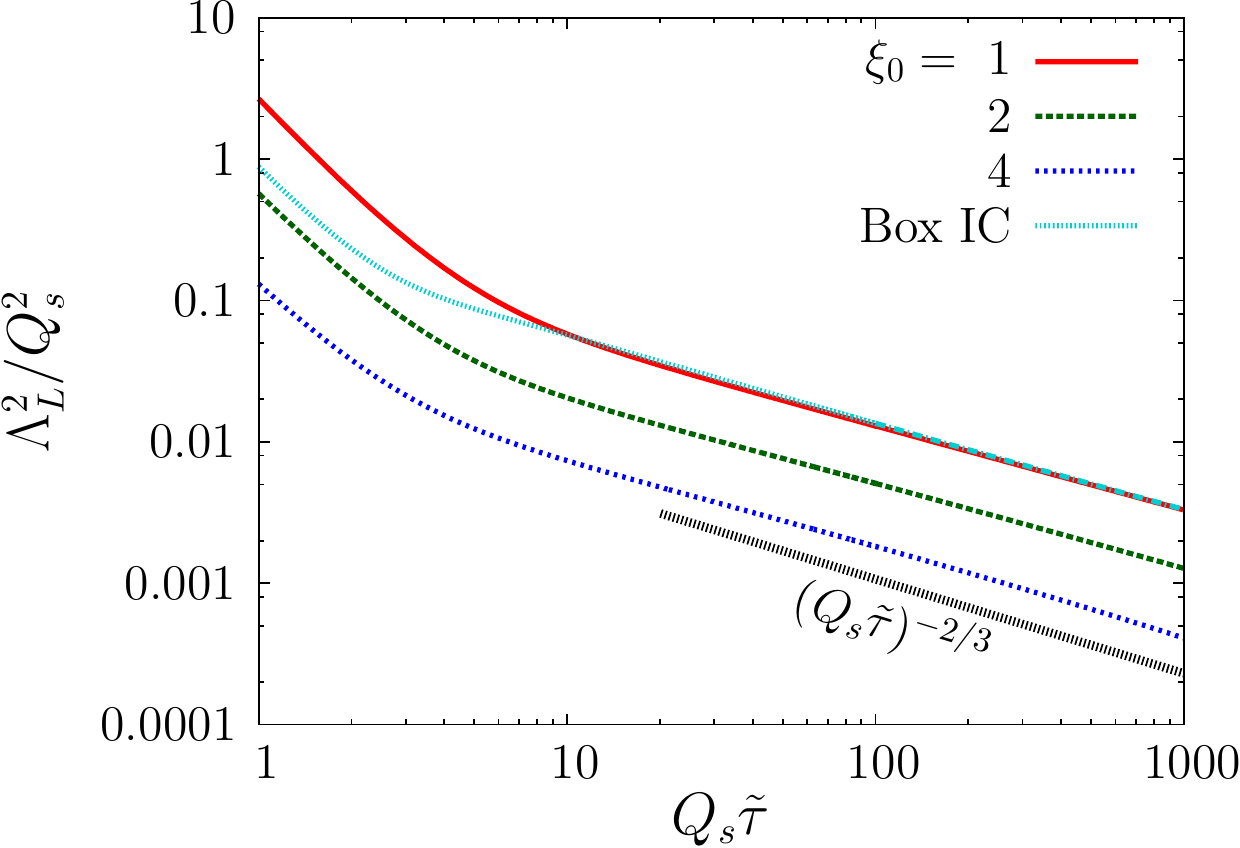} \vspace{-10pt}
  \caption{Time evolution of the longitudinal hard scale. 
   Several values of the anisotropy parameter $\xi_0$ are compared for the initial distribution in Eq.~\eqref{ini_dist}.
   Box IC denotes the initial distribution given by Eq.~\eqref{boxIC}.}
  \label{fig:hs_g3}
 \end{center}
\end{figure}

The ratio of the longitudinal pressure to the transverse pressure is a measure of the bulk anisotropy of the system. 
As shown in Fig.~\ref{fig:press_g1}, the ratio decreases in time because of the longitudinal expansion of the system.
For larger couplings, it turns to increase at later times and approaches unity. 
For weak couplings, the system falls into the scaling regime. 
If the pressure is dominated by hard modes, this ratio should show the temporal power law behavior $\sim (Q_s \tau )^{-2/3}$ in the scaling regime. 
However, as seen in Fig.~\ref{fig:press_g1}, the curves clearly deviate from the scaling behavior $(Q_s \tau )^{-2/3}$ -- more significantly than that observed for the longitudinal hard scale. The same tendency was noted in the classical-statistical simulations; this is because the longitudinal pressure is not dominated by the hard modes and  soft modes contribute significantly \cite{Berges:2015ixa}.

To check the robustness of the classical scaling behavior, we compare different initial distributions for the longitudinal hard scale in Fig.~\ref{fig:hs_g3}. 
For the initial distributions given by Eq.~\eqref{ini_dist}, we compute results for anisotropy parameter values of $\xi_0=1$, 2 and 4.
Furthermore, results for a different functional form of the initial distribution,
\begin{equation} \label{boxIC}
f(\tau_0 ,\pperp ,p_z )= \frac{n_0}{g^2} \theta (Q_s -\sqrt{\pperp^2 +p_z^2}) \, ,
\end{equation}  
denoted as Box IC, are compared to the results from distributions with the initial conditions in Eq.~\eqref{ini_dist}. 
The other parameters in the computation are set to $g=10^{-2}$, $n_0=0.1$ and $Q_s \tilde{\tau}_0=1$. 
For all initial distributions, the longitudinal hard scale shows an identical classical scaling behavior.

\subsection{Effects of the time dependence of the Coulomb logarithm} \label{subsec:CL}
In some of the earlier studies that employ the kinetic equation in the small angle approximation, the Coulomb logarithm \eqref{cl} is taken to be a time independent constant \cite{Blaizot:2013lga,Blaizot:2014jna}. In the present study, since the Debye mass given by Eq.~\eqref{Debye} is a time dependent IR scale, the Coulomb logarithm varies in time. We  will investigate the effect of this time dependence in this subsection. 

\begin{figure}[tb]
 \begin{center}
  \includegraphics[clip,width=7.3cm]{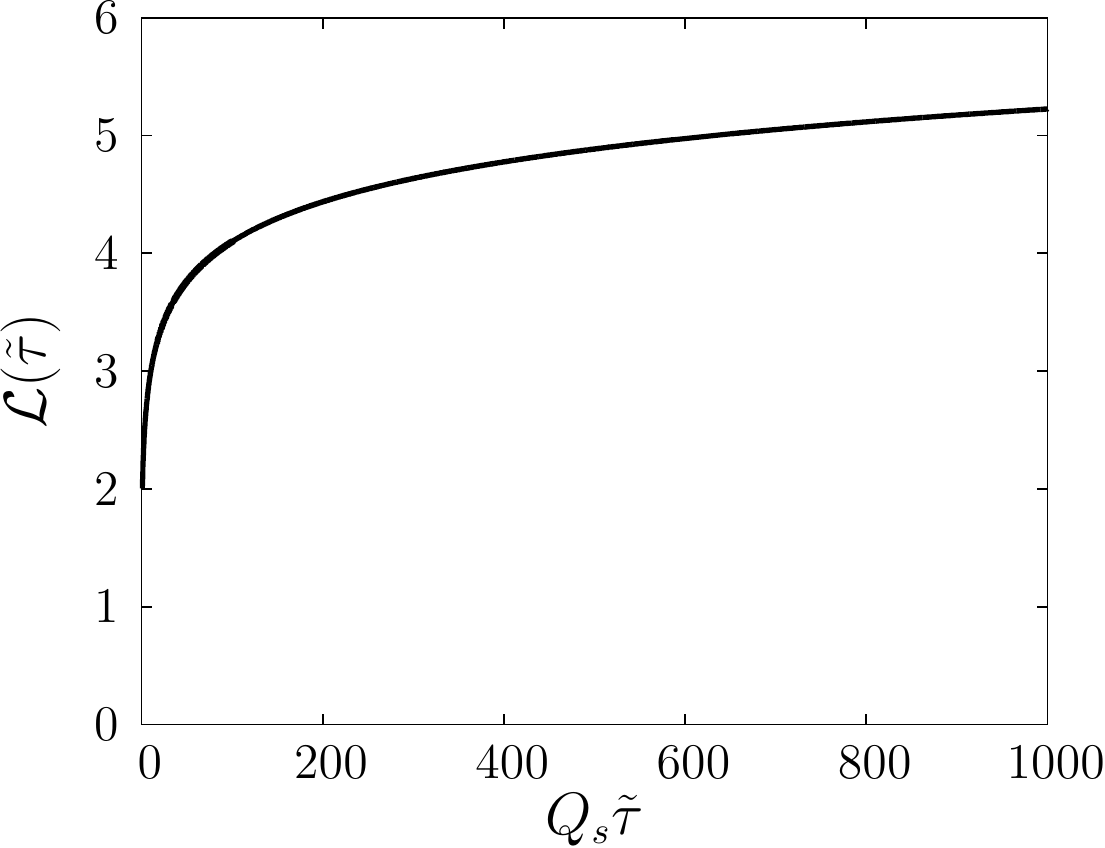} \vspace{-10pt}
  \caption{Time evolution of the Coulomb logarithm.}
  \label{fig:cl}
 \end{center}
\end{figure}

\begin{figure}[tb]
 \begin{center}
  \includegraphics[clip,width=8cm]{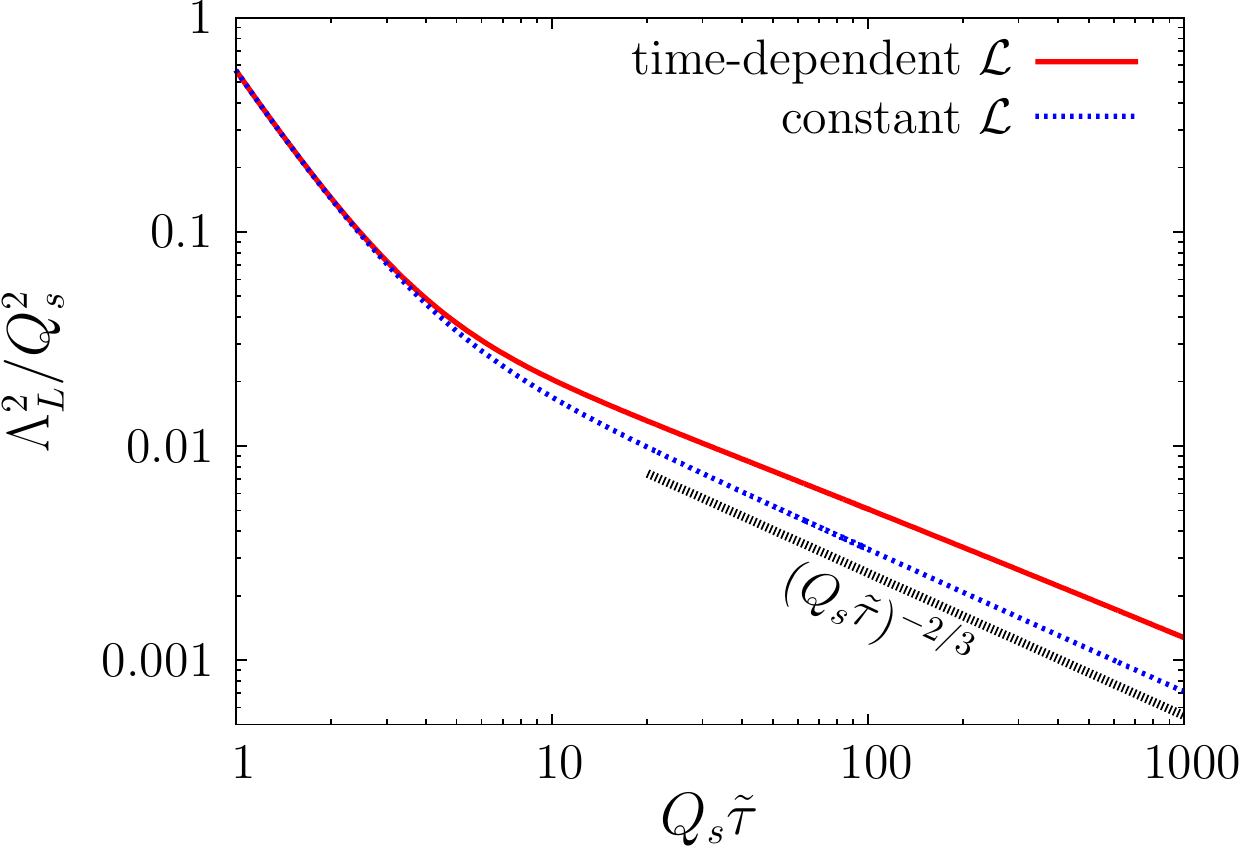} \vspace{-10pt}
  \caption{The time evolution of the longitudinal hard scale. 
  The result for the case where the time dependence of the Coulomb logarithm is taken into account is compared to the result 
   where logarithm is fixed to its initial value.}
  \label{fig:hs_g4}
 \end{center}
\end{figure}

This is demonstrated in Fig.~\ref{fig:cl} for the parameters $g=10^{-2}$, $n_0=0.1$, $\xi_0=2$ and $Q_s \tilde{\tau}_0=1$. 
As the Debye mass decreases in time as $m_D^2 \sim 1/\tau$, the Coulomb logarithm increases in time logarithmically.
The increase is slow except for early times. Therefore the classical scaling behavior is not spoiled by the time dependence of the IR cutoff.
The only effect is the change of the effective interaction strength.  Because $\mathcal{L}$ appears as an overall multiplicative factor in the collision terms, the increase of $\mathcal{L}$ indicates that the time evolution is faster. 
This can be confirmed in Fig.~\ref{fig:hs_g4}, where the longitudinal hard scale is plotted for the case where the time dependence of $\mathcal{L}$ is taken into account and compared to results where $\mathcal{L}$ is fixed to its initial value\footnote{In the computation of the latter case, we have used $N_\kappa=512$ for the number of angular grid points in order to precisely calculate $\Lambda_L^2$ at later times.}. The transition from the free streaming regime to the classical scaling regime happens later in the constant $\mathcal{L}$ computation, indicating the effective interaction strength is weaker in this case.

\section{Numerical results II. $N_f=3$} \label{sec:quark_results}
Now that we have confirmed that the kinetic equation within the small angle approximation can qualitatively describe the classical scaling behavior of overpopulated gauge fields, we will add $N_f=3$ quark degrees of freedom to the expanding Glasma. We will assume that the initial distribution for quarks is given by 
\begin{equation}
F (\tau_0 ,p_T,p_z)= F_0 \, e^{-\left[ \pperp^2 +(\xi_0 p_z )^2 \right]/Q_s^2} \, .
\end{equation}
The shape of this initial distribution is irrelevant for the following discussion on the classical scaling behavior; for simplicity, we have therefore chosen the same momentum dependence as that for the initial gluon distribution. We note that on account of the Pauli principle, the initial occupancy $F_0$ cannot exceed unity. 

\subsection{Self-similar evolution} \label{subsec:quark_results}
\begin{figure}[tb]
 \begin{center}
  \includegraphics[clip,width=8cm]{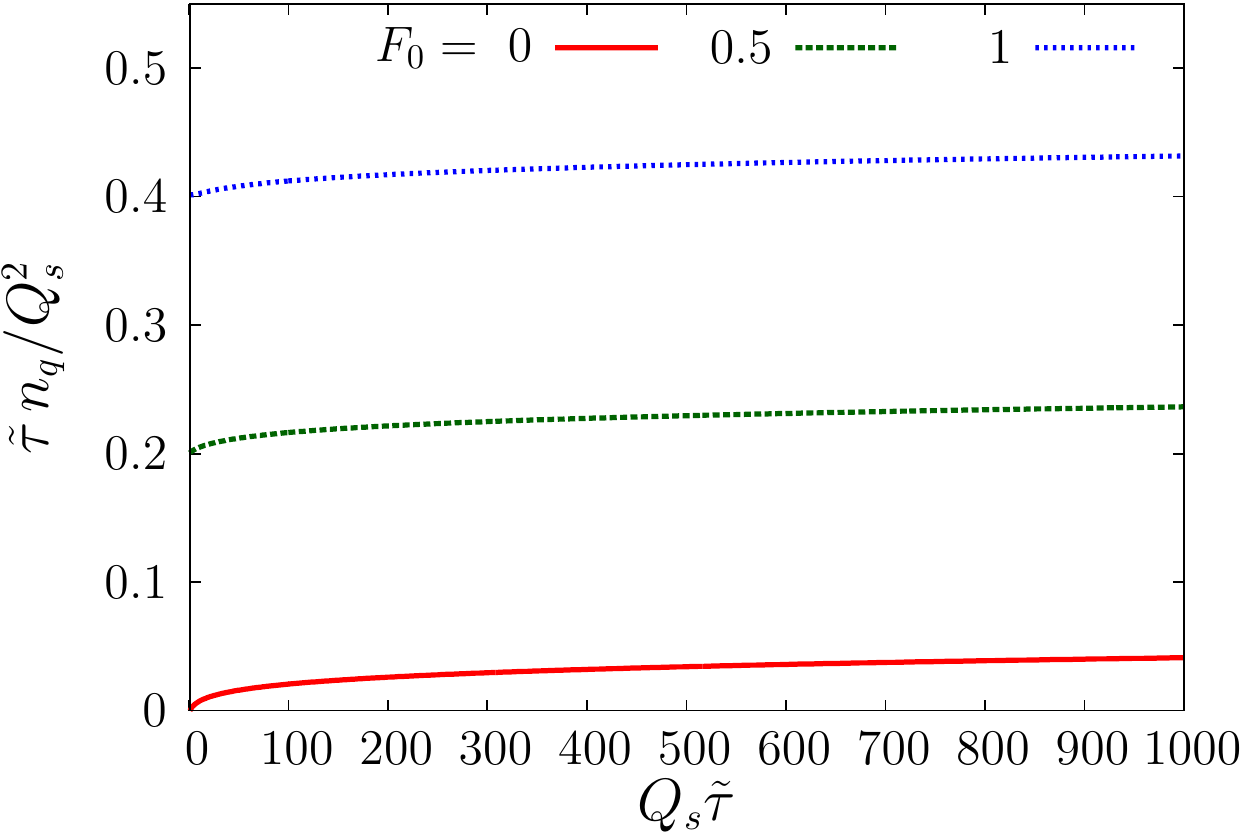} \vspace{-10pt}
  \caption{The time evolution of the quark number density multiplied by time. Different values of the quark initial occupancy $F_0$ are compared. }
  \label{fig:numq}
 \end{center}
\end{figure}

Figure \ref{fig:numq} displays the time evolution of the quark number density multiplied by time. 
Three different values for the initial quark occupancies $F_0$ are compared. 
The parameters used in these computations are $g=10^{-2}$, $n_0=0.1$, $\xi_0=2$, and $Q_s \tilde{\tau}_0=1$. 
If quark number is conserved, $\tilde{\tau} n_q$ should stay constant. 
The increase in this quantity with time, seen in Fig.~\ref{fig:numq}, clearly indicates the effect due to quark pair production via gluon fusion. 
The contribution from this process is however rather mild because the effective interaction strength with gluons is governed by the factor $g^2 f_0=n_0$; as our choice of $n_0$ indicates, this 
quantity is small in the regime where the kinetic description is valid. 
On the other hand, at early times in heavy-ion collisions, $\tau \ltsim Q_s^{-1}$, the factor $g^2 f_0$ is of order unity and quark pair production can be quite large due to nonperturbative processes, as hinted at by extant real-time lattice simulations~\cite{Gelis:2005pb,Gelfand:2016prm,Tanji:2016dka}.

\begin{figure}[tb]
\begin{tabular}{cc}
 \begin{minipage}{0.5\hsize}
  \begin{center}
   \includegraphics[clip,width=8cm]{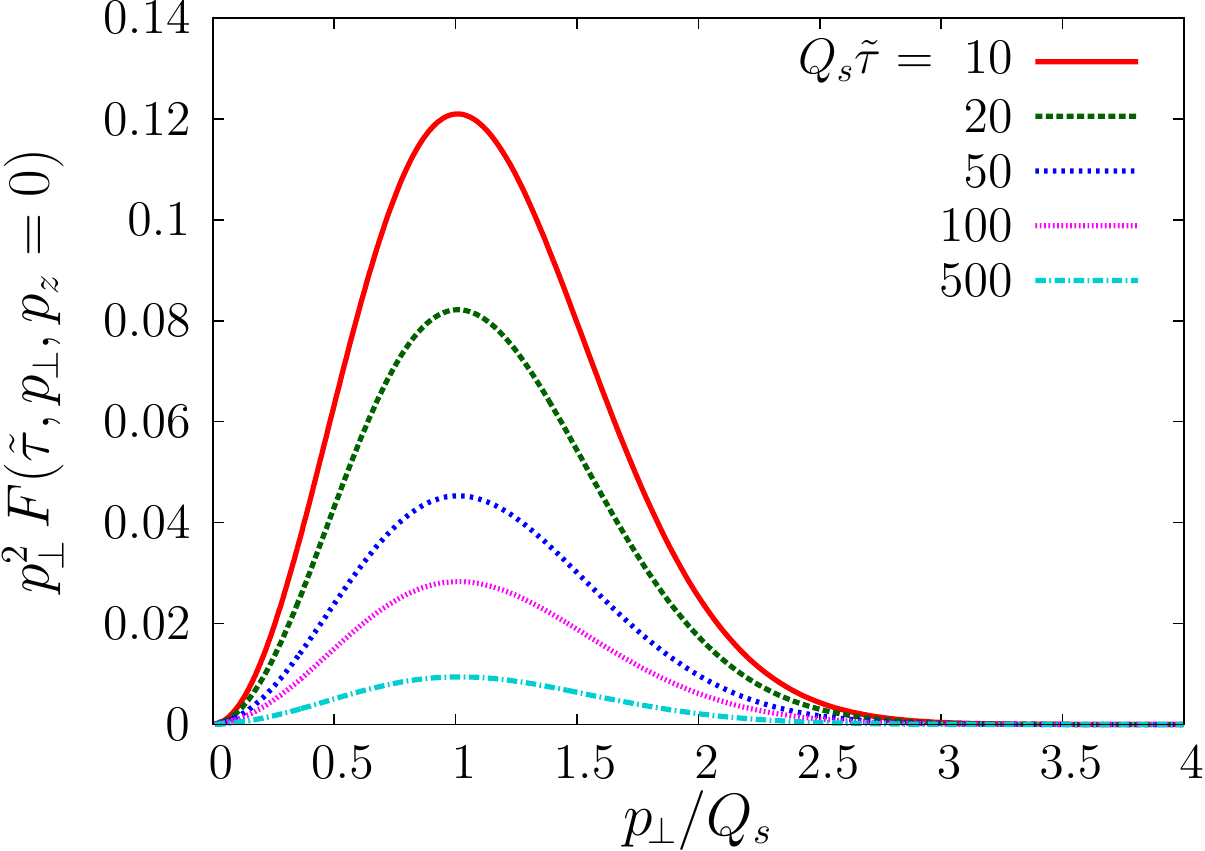} 
  \end{center}
 \end{minipage} &
 \begin{minipage}{0.5\hsize}
  \begin{center}
   \includegraphics[clip,width=8cm]{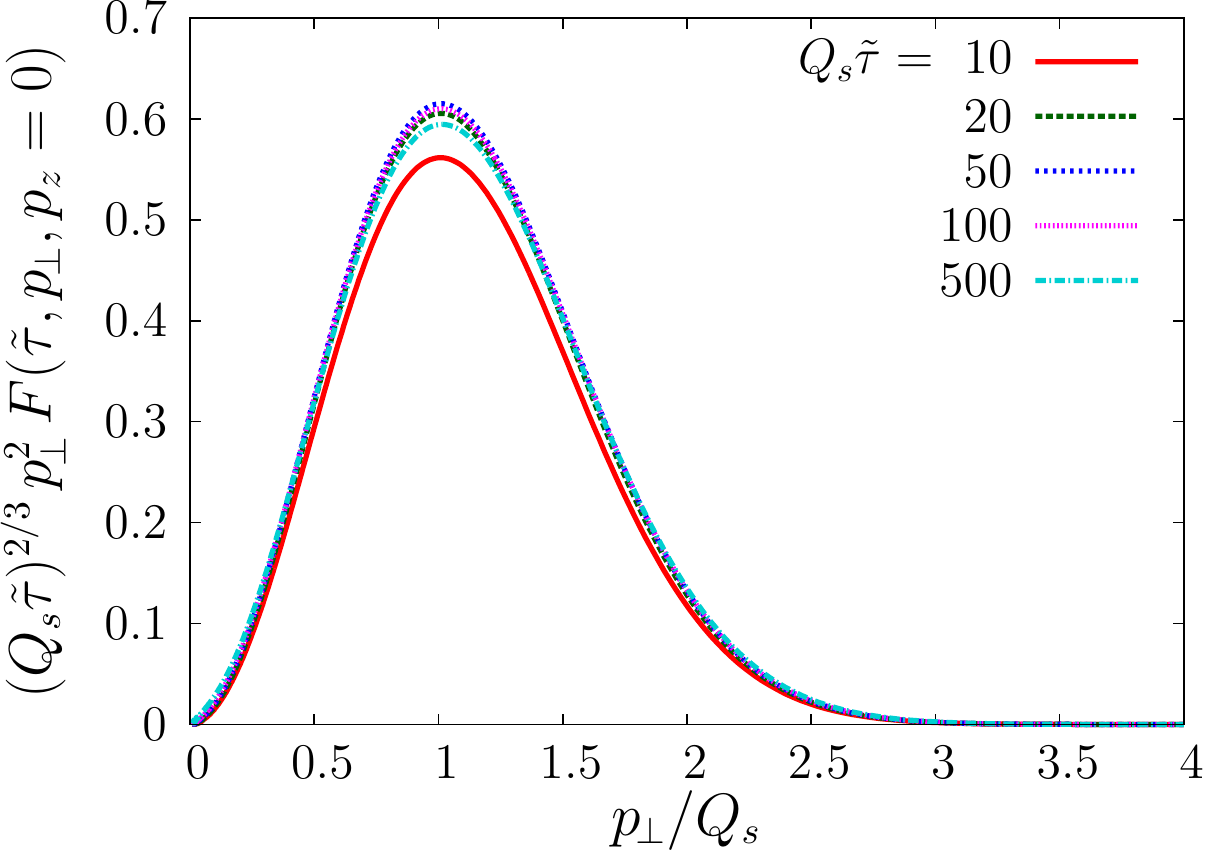} 
  \end{center}
 \end{minipage} 
\end{tabular}
\caption{The second moment of the transverse momentum distribution for quarks, $\pperp^2 F (\tilde{\tau}, \pperp ,0)$, for  different times.
Left panel: original distribution. Right panel: rescaled distribution.
The typical transverse momentum scale remains centered at $\sim Q_s$. }
\label{fig:distTq2}
\end{figure}

The second moment of the quark distribution evaluated at $p_z=0$ is plotted in the left panel of Fig.~\ref{fig:distTq2} for different times. 
The initial quark occupancy is chosen to be $F_0=0.5$, and the other parameters are the same as those for Fig.~\ref{fig:numq}. 
Here we show only the quark distribution because the gluon distribution is little affected by the existence of quarks for Glasma initial conditions.  
We observe that, just as in the gluon case, the typical transverse momentum is centered around $Q_s$.
In the right panel of Fig.~\ref{fig:distTq2}, we plot the quark distribution multiplied by the factor $(Q_s \tilde{\tau})^{2/3}$. 
At later times, $Q_s \tilde{\tau} \gtsim 20$, the rescaled distributions at different times overlap nicely.  
This indicates that the quark transverse momentum distribution obeys the same scaling law as the gluon transverse momentum distribution.

\begin{figure}[tb]
\begin{tabular}{cc}
 \begin{minipage}{0.5\hsize}
  \begin{center}
   \includegraphics[clip,width=8cm]{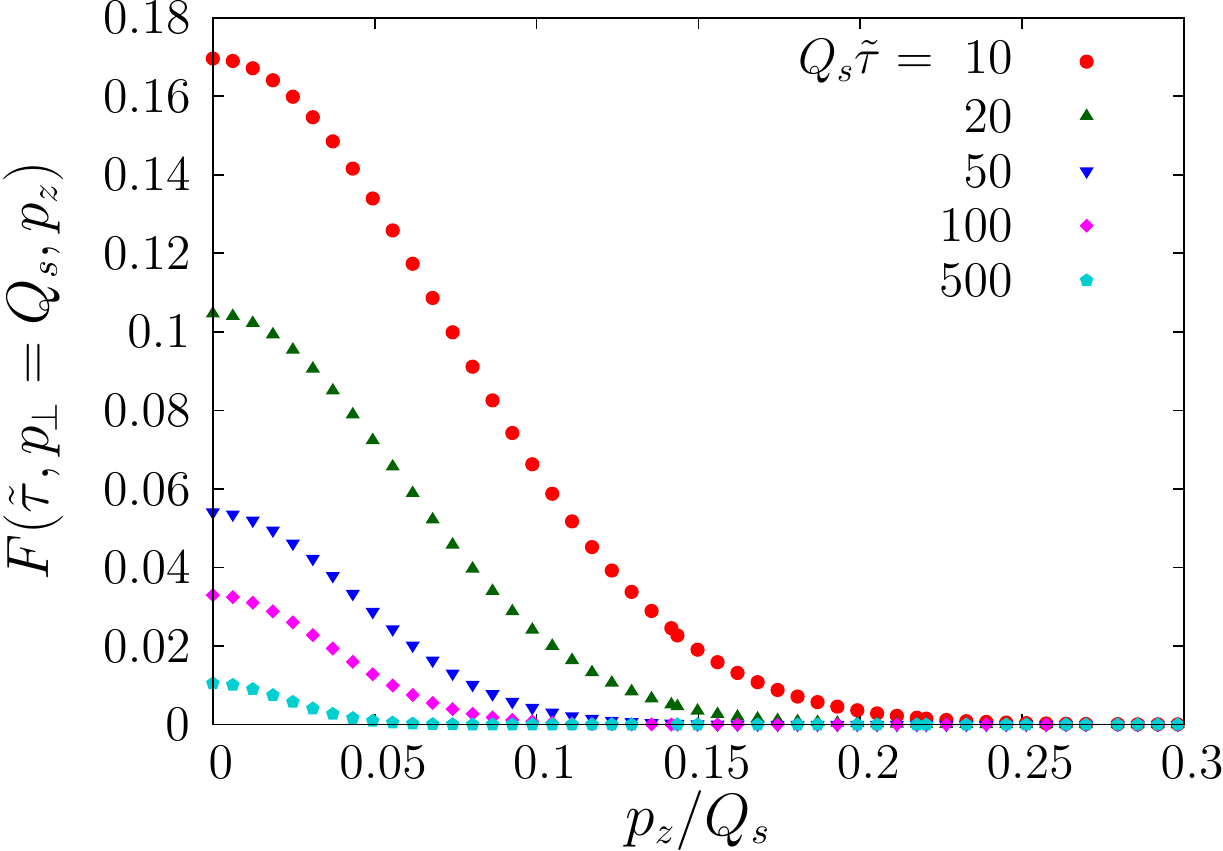}
  \end{center}
 \end{minipage} &
 \begin{minipage}{0.5\hsize}
  \begin{center}
   \includegraphics[clip,width=8cm]{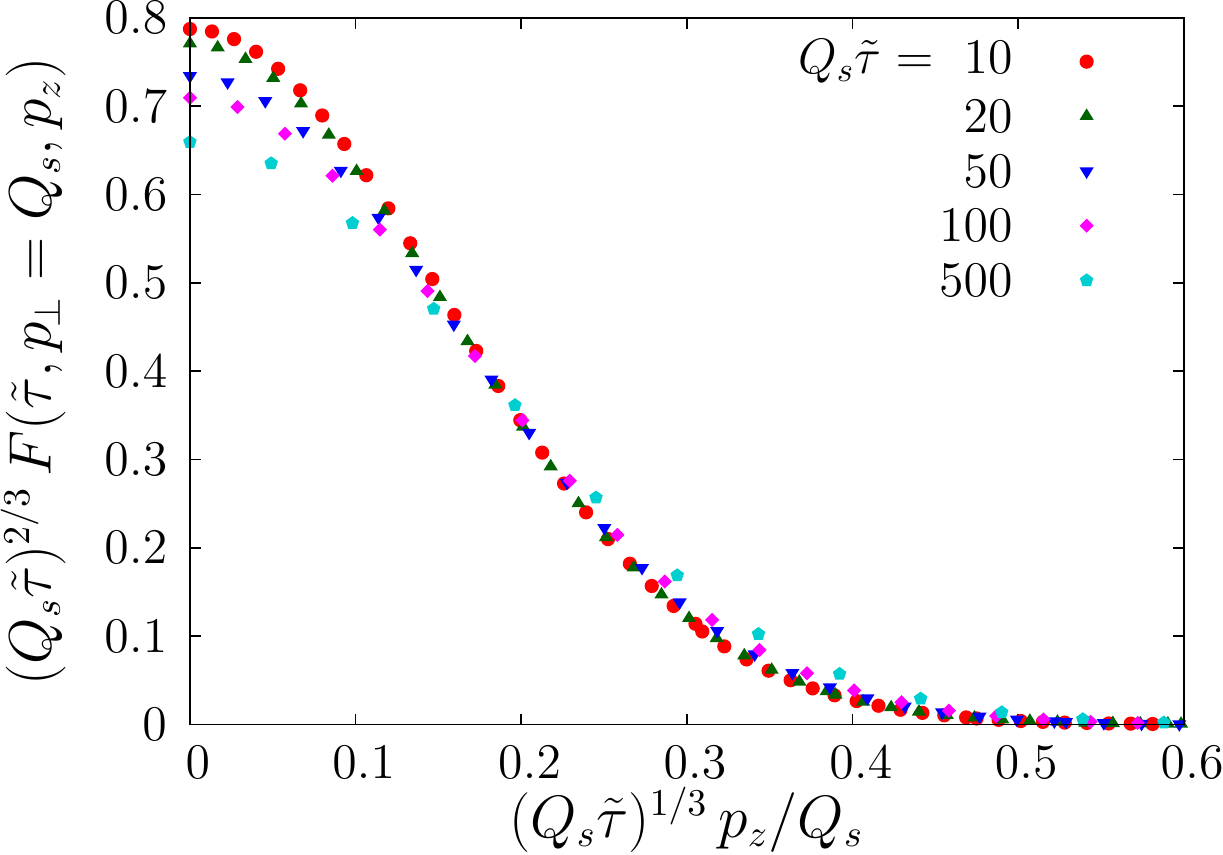}
  \end{center}
 \end{minipage} 
\end{tabular}
\caption{The time evolution of the longitudinal momentum distribution for quarks evaluated at $\pperp =Q_s$.
Left panel: original distribution. Right panel: rescaled distribution. }
\label{fig:distLq1}
\end{figure}

\begin{figure}[tb]
 \begin{center}
  \includegraphics[clip,width=8cm]{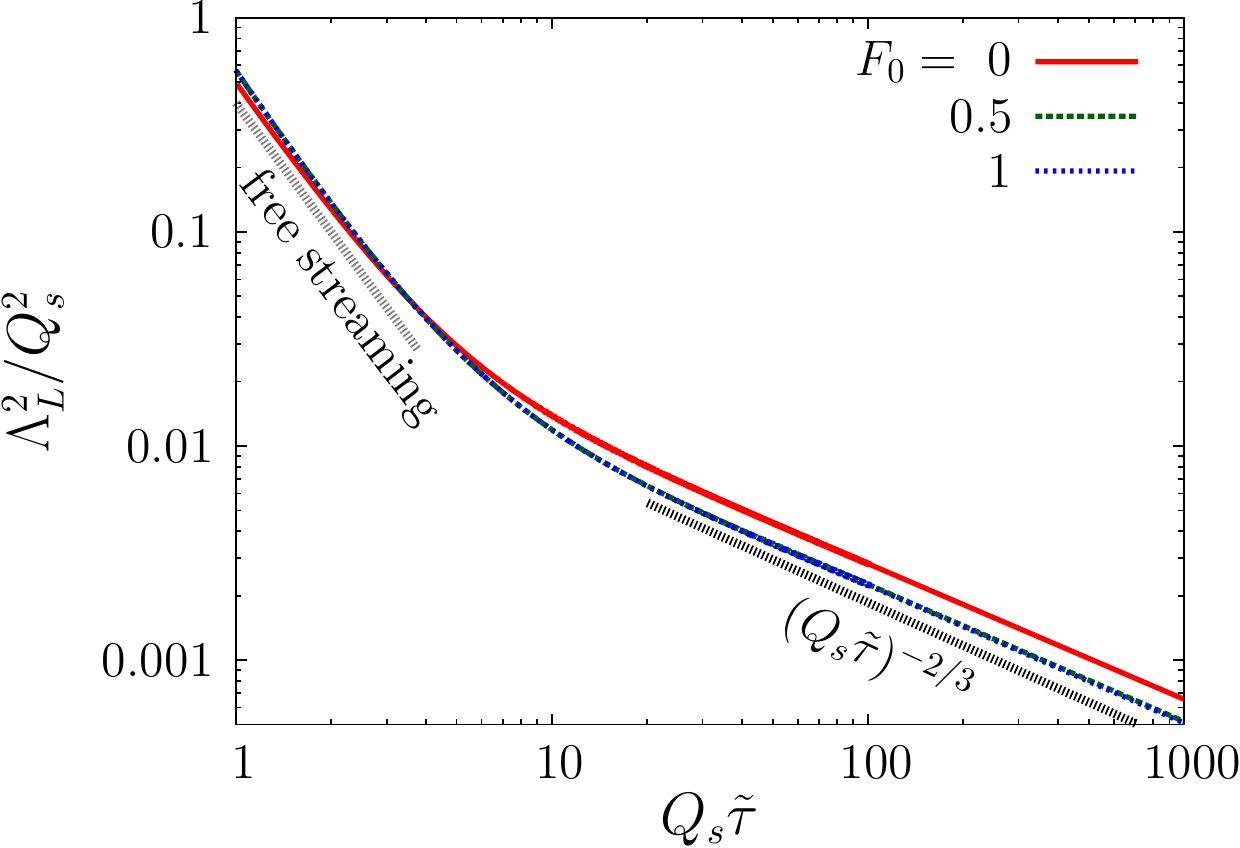} \vspace{-10pt}
  \caption{Time evolution of the longitudinal hard scale for quarks. 
   Three different values of the initial quark occupancy $F_0$ are compared. The results for $F_0=0.5$ and $F_0=1$ are indistinguishable.}
  \label{fig:hs_q1}
 \end{center}
\end{figure}

It is not just the transverse momentum distributions that satisfy scaling laws. The quark longitudinal momentum distributions satisfy the same scaling law as that for gluons. This is displayed in Fig.~\ref{fig:distLq1}. These observations can be summarized into the scaling expression,
\begin{equation} \label{scalingQ}
F(\tau ,\pperp ,p_z ) = (Q_s \tau )^{-2/3} F_S \left( \pperp ,(Q_s \tau )^{1/3} p_z \right) \, ,
\end{equation}
which has the identical form as that of Eq.~\eqref{scaling} for gluons. 
This scaling behavior is further confirmed in Fig.~\ref{fig:hs_q1}, where the time evolution of the longitudinal hard scale is plotted for quark distributions with different values of the initial occupancy $F_0$. We observe that a transition from free streaming behavior to the scaling form occurs for any value of $F_0$.

At first sight, the exhibition of scaling behavior by the quark distribution might appear strange because  Bose enhancement is a  prerequisite for the emergence of the scaling behavior.
However this is in fact not at all strange because what needs to be Bose-enhanced is a scattering amplitude.
When quarks elastically scatter with gluons, as is transparent from the kinetic equations, the presence of gluons in the final state causes the process to  be Bose-enhanced. 
In both the equation for gluons \eqref{kine_g1},  and that for quarks \eqref{kine_q1}, the diffusion terms involve the common factor $I_a$, which can be interpreted as the density of scatterers. The integral $I_a$ in the Glasma is dominated by the gluon distribution, and it is therefore enhanced by the Bose factor. 
Thus quarks undergo small angle scattering processes that are Bose-enhanced just as for gluons, and therefore exhibit the same scaling behavior. 

To demonstrate the emergence of scaling behavior for the quark distributions more explicitly, we will follow the argument for gluons presented in Ref.~\cite{Berges:2013fga}.   
In the classical scaling regime of the Glasma, the typical longitudinal momentum is much smaller than the typical transverse momentum.
Therefore, the collision term in the small angle approximation is dominated by the derivative term in $p_z$. 
Furthermore, as observed in Fig.~\ref{fig:numq}, quark pair production is slow in the kinetic regime, and the source term may be neglected. 
The kinetic equation for quarks (Eq.~\eqref{kine_q1}) can therefore be approximated by 
\begin{equation} \label{kine_q_simp}
\left( \frac{\partial}{\partial \tau} -\frac{p_z}{\tau} \frac{\partial}{\partial p_z} \right) F(\tau ,\pperp ,p_z ) 
= \hat{q} \frac{\partial^2}{\partial p_z^2} F(\tau ,\pperp ,p_z ) \, ,
\end{equation}
where  we define $\hat{q}$ to be 
\begin{equation}
\hat{q} = \frac{g^2}{4\pi} N_c C_F \mathcal{L} \int \! \frac{d^3 p}{(2\pi)^3} \, f^2 \, ,
\end{equation}
a quantity proportional to the integral $I_a$ and dominated by the gluon distribution.
Since the gluon distribution has the scaling form specified in  Eq.~\eqref{scaling}, the time dependence for $\hat{q}$ can be identified as 
\begin{equation}
\hat{q} \sim (Q_s \tau )^{-5/3} 
\end{equation}
up to the logarithmic time dependence contained in $\mathcal{L}$. 
If we assume that the following scaling form of the quark distribution
\begin{equation}
F( \tau ,\pperp ,p_z ) = (Q_s \tau )^{\alpha^\prime} 
F_S \left( (Q_s \tau)^{\beta^\prime} \pperp ,(Q_s \tau)^{\gamma^\prime} p_z \right)
\end{equation}
is a stationary solution of the simplified kinetic equation \eqref{kine_q_simp}, it follows that
\begin{equation}
\gamma^\prime = \frac{1}{3} \, .
\end{equation}
Furthermore, from the approximate number conservation for quarks, $n_q \sim 1/\tau$, one obtains a constraint
\begin{equation}
\alpha^\prime -2\beta^\prime -\gamma^\prime = -1 \, .
\end{equation}
Since the typical longitudinal momentum is much smaller than the typical transverse momentum, the longitudinal pressure density $\mathcal{P}_L$ is negligible compared with the energy density $\mathcal{E}$. 
In this case, the energy density approximately behaves as $\mathcal{E} \sim 1/\tau$,
from which it follows that
\begin{equation}
\alpha^\prime -3\beta^\prime -\gamma^\prime = -1 \, .
\end{equation} 
By these, the remaining exponents are determined to be 
\begin{equation}
\alpha^\prime = -\frac{2}{3} \, , \hspace{10pt}
\beta^\prime = 0 \, .
\end{equation}
To summarize, quark distributions show the same scaling behavior as that exhibited by the gluon distribution when (i) the diffusion constant $I_a$ is dominated by overoccupied gluons, (ii) the typical longitudinal momentum is much smaller than the typical transverse momentum, and (iii) the quark number density $\tau n_q$ does not vary rapidly in time. 

\begin{figure}[tb]
 \begin{center}
  \includegraphics[clip,width=8cm]{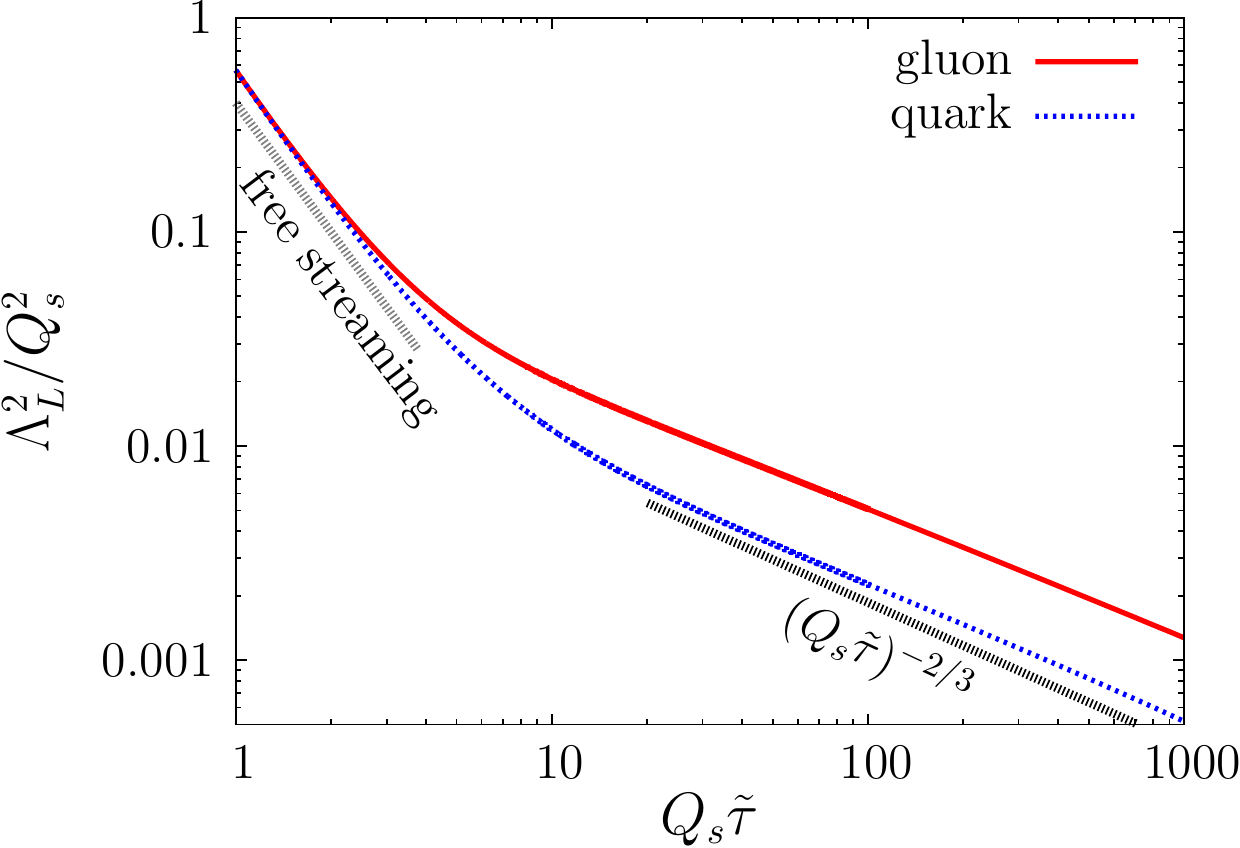} \vspace{-10pt}
  \caption{Time evolution of the longitudinal hard scale. 
  A comparison of the scales for gluons and that for quarks is shown.}
  \label{fig:hs_q2}
 \end{center}
\end{figure}

\begin{figure}[tb]
 \begin{center}
  \includegraphics[clip,width=8cm]{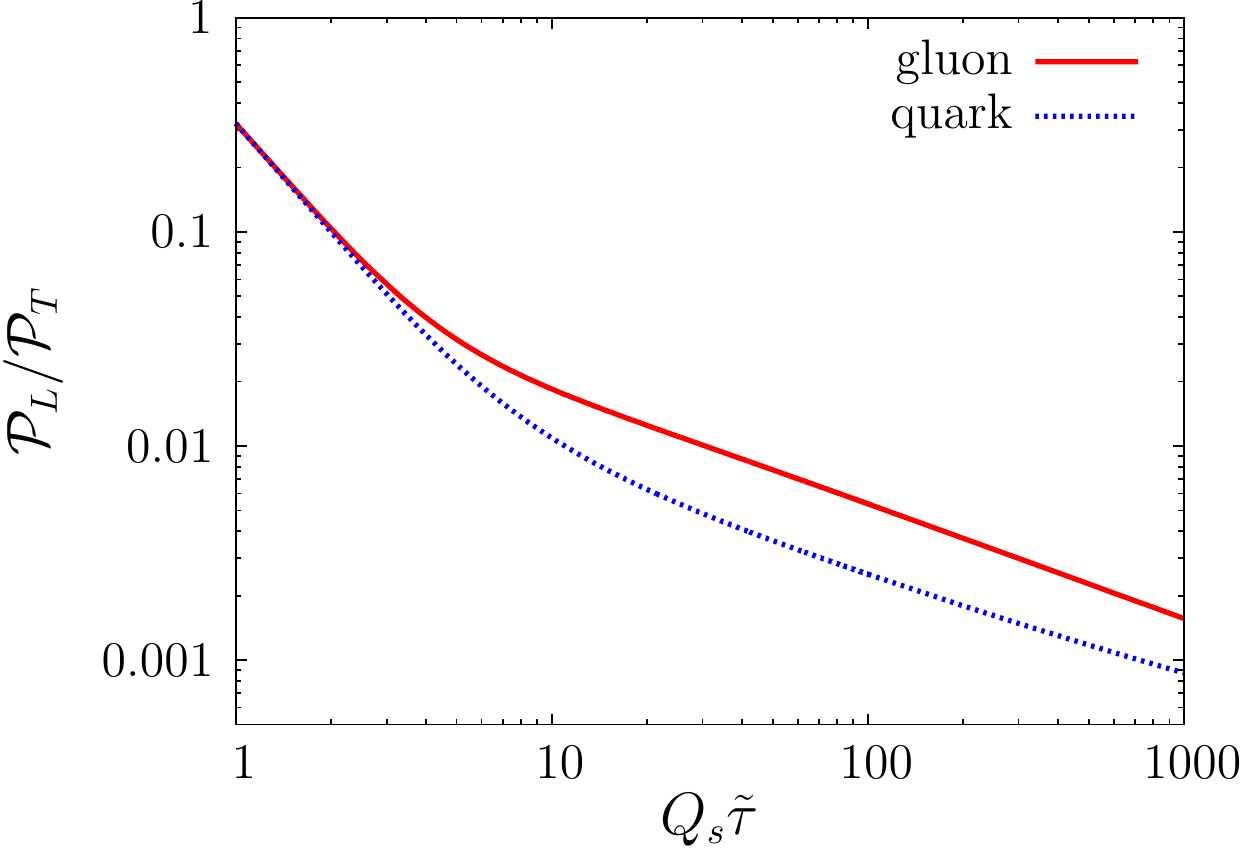} \vspace{-10pt}
  \caption{Time evolution of the pressure ratio for gluons, $\mathcal{P}_{L,g}/\mathcal{P}_{T,g}$, and that for quarks, $\mathcal{P}_{L,q}/\mathcal{P}_{T,q}$.}
  \label{fig:press_q1}
 \end{center}
\end{figure}

Though quark distributions qualitatively obey the same scaling behavior as those for gluons, there are quantitative differences in their temporal evolution. 
In Fig.~\ref{fig:hs_q2}, the longitudinal hard scale is compared for gluons and quarks ($F_0=0.5$). 
The transition from the free streaming behavior to the classical scaling behavior happens later for quarks than gluons. 
This can be understood simply as a consequence of the smaller effective coupling of quarks seen in the comparison between Eqs.~\eqref{kine_g1} and \eqref{kine_q1}. 
A similar tendency is observed in the bulk anisotropy (the ratio of the longitudinal pressure to the transverse pressure) plotted in Fig.~\ref{fig:press_q1}. 
The quark distribution is more anisotropic relative to the gluon distribution as quarks scatter more weakly than gluons.

\subsection{Approach to equilibrium} \label{subsec:equilibrium}
In the bottom-up thermalization scenario, the inelastic collinear splitting process that produces soft gluons plays a crucial role in thermalization. 
As the kinetic equations \eqref{kine_g1} and \eqref{kine_q1} contain only elastic collision terms, they are applicable only to the classical scaling regime in bottom-up thermalization. 

Nevertheless, albeit not justified on parametric grounds, it is an amusing exercise to crank up the coupling and see how both kinetic and chemical equilibration occur via only elastic collisions. 
This is for instance the strategy adopted in a number of phenomenological models of heavy-ion collisions -- and it is useful to understand, from this perspective at least, how equilibration occurs. 
With only elastic collisions included ,  thermalization takes a parametrically long time $\sim Q_s^{-1} \exp (\alpha_s^{-1/2})$ \cite{Mueller:1999fp,Bjoraker:2000cf,Serreau:2001xq} -- much longer than that in the bottom-up scenario.

In order to observe the approach to equilibrium in a practical computational time, we take a relatively large value for the coupling $g=0.5$ and the gluon occupation coefficient $n_0=3/4$, where admittedly the validity of the kinetic equations is marginal. 
Furthermore, we use the initial anisotropy parameter $\xi_0=1$ and a relatively large initial time $Q_s \tilde{\tau}_0 =100$.\footnote{%
These conditions correspond to the initial time $Q_s \tau_0 \simeq 100$ in terms of the original time variable.}
For the quark initial occupation number, we compare different values $F_0=0$, 0.5 and 1. 
When the system gets close to equilibrium, the assumption of a constant UV cutoff $q_\text{max}=Q_s$ in the Coulomb logarithm is inadequate;  
we will instead use the mean transverse momentum in Eq.~\eqref{mean_pT} as the cutoff. 

\begin{figure}[tb]
 \begin{center}
  \includegraphics[clip,width=8.5cm]{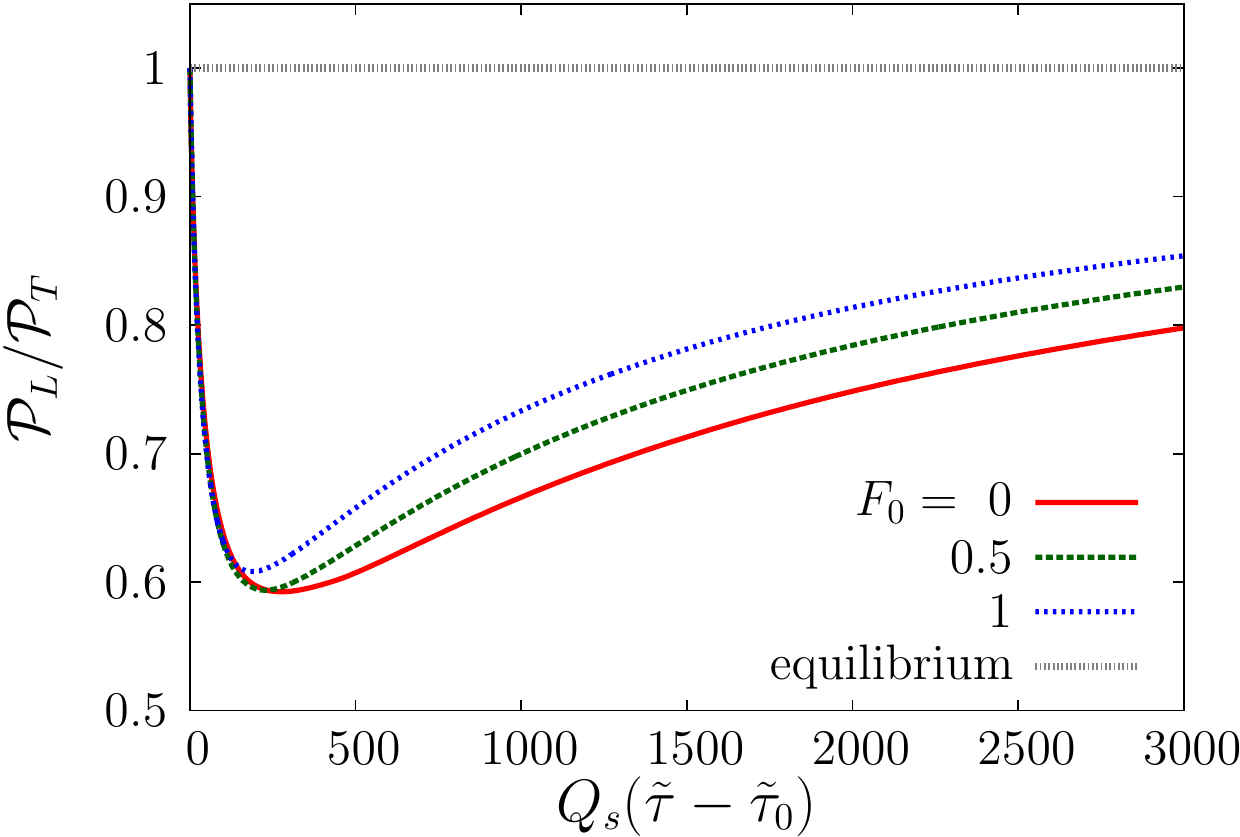} \vspace{-10pt}
  \caption{The ratio of the longitudinal pressure to the transverse pressure as a function of time for the coupling $g=0.5$.
  Three different values of the initial quark occupancy are compared.}
  \label{fig:press_q2}
 \end{center}
\end{figure}

The ratio of pressures, $\mathcal{P}_L /\mathcal{P}_T$, is plotted as a function of time in Fig.~\ref{fig:press_q2}. 
The longitudinal pressure $\mathcal{P}_L$ and the transverse pressure $\mathcal{P}_T$ include here the contributions from both gluons and quarks.
The ratio decreases at early times due to the longitudinal expansion--unlike the result for weaker couplings however, already at $g=0.5$ one observes a turnaround in this ratio and the 
ratio begins to increase. While full isotropization is not achieved\footnote{It should be noted though that the value of the coupling, $g=0.5$, that we choose, $Q_s$ is enormous and is far above the TeV scale -- hence values 
of $Q_s\tau\sim 1000$ on plots still correspond to times far less than a Fermi. The fact that  the magnitude $Q_s$ should be commensurate with the value of $g$ chosen is often obscured in a number of works, where $Q_s\sim 1$ GeV is assumed even for very weak couplings.} by $Q_s (\tilde{\tau}-\tilde{\tau}_0)=3000$, the ratio reaches 0.8. 

\begin{figure}[tb]
 \begin{center}
  \includegraphics[clip,width=8.5cm]{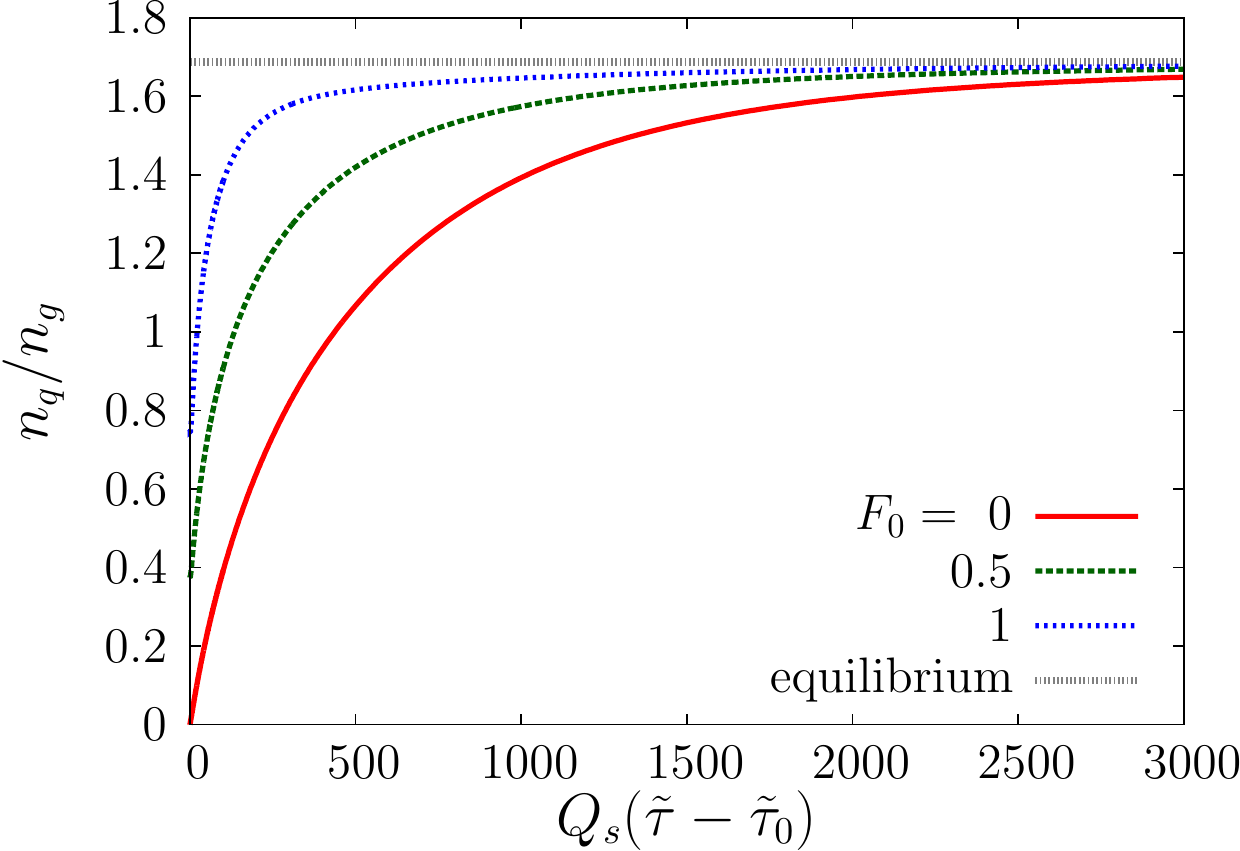} \vspace{-10pt}
  \caption{The ratio of the quark number and the gluon number as a function of time for the coupling $g=0.5$.
  Three different values of the quark initial occupancy are compared.
  The black dashed line denotes the value expected in the thermal equilibrium.}
  \label{fig:nratio1}
 \end{center}
\end{figure}

In Fig.~\ref{fig:nratio1}, the ratio of the quark number and the gluon number is plotted as a function of time. 
In thermal equilibrium with vanishing chemical potential, the ratio should be $3N_c N_f/(2(N_c^2-1))=27/16$, which is shown as a gray dashed line. 
The ratio of quark and gluon densities approaches its equilibrium value more quickly relative to the ratio of pressures.
The simple message is that while pressure isotropization requires the system to overcome the effect of expansion, in contrast, equipartition is unhindered since both quarks and gluons experience the expansion.

\section{Summary and outlook} \label{sec:summary}

We presented in this paper numerical results from the solution of kinetic equations for an expanding Glasma of overoccupied gluons and quarks. We considered only elastic \ttt scatterings amongst the quarks and gluons. As may be anticipated, we observed that a cascade of particle number to the infrared develops generating an enhancement in this region; this phenomenon can be interpreted heuristically as signaling the formation of a transient Bose-Einstein condensate. For $N_f=0$, and for small values of the coupling where classical assumptions are robust, we find that our simple kinetic equation for the Glasma qualitatively reproduces key features of classical-statistical real-time lattice simulations of the 3+1-dimensional Yang-Mills equations. These numerical simulations, for sufficiently small couplings, should completely capture the physics of the expanding Glasma. 

The fact that results of our kinetic simulations reproduce the key features of the classical-statistical simulation results is truly remarkable. This is because, on the face of it, these kinetic equations do not include $2\leftrightarrow 3$ scatterings that are also present in the classical stage of bottom-up thermalization. Such inelastic scatterings would,  in principle, prevent a BEC from developing~\cite{Kurkela:2011ti,York:2014wja,Blaizot:2016iir}. Since a first principles weak coupling {\it kinetic} treatment should contain these contributions, the bottom-up kinetic scenario does not include the presence of a transient BEC. However the bottom-up framework is also not fully robust because, as we noted previously, it does not contain the effects of late time plasma instabilities that would modify the temporal evolution of the Glasma~\cite{Kurkela:2011ti}. As discussed in detail in \cite{Berges:2015ixa}, the fact that the bottom-up scenario captured some of the key features of the numerical simulations posed a challenge for conventional {\it ab initio} kinetic treatments of the Glasma. 

Further, as also noted in \cite{Berges:2015ixa}, the bottom-up framework was unable to reproduce the temporal behavior of the anisotropy measure $P_L/P_T$ seen in the numerical simulations. From the bottom-up based simulations presented in the literature~\cite{Kurkela:2015qoa}, it is unclear whether the $1/\pperp$ dependence of the single particle distributions is reproduced. The fact that our simplified kinetic treatment with only \ttt scattering captures qualitative features of the overoccupied Glasma therefore confirms the conjecture in \cite{Berges:2015ixa} that the infrared dynamics is playing a bigger role than anticipated not only in suppressing late time plasma instabilities but in the evolution of the bulk anisotropy as well. 

The possibility that a transient BEC  might provide an effective description of infrared dynamics in the Glasma was conjectured in \cite{Blaizot:2011xf} and developed further in \cite{Blaizot:2013lga}. However these estimates were made without the benefit of detailed kinetic simulations of the {\it expanding} Glasma and assumed a constant value of $P_L/P_T$ even for weak couplings. This constant value for the anisotropy was disfavored by classical-statistical simulations. It is therefore perhaps ironic that an effective BEC description of the classical stage of the Glasma survives in a qualitative comparison of our effective kinetic description and the full classical-statistical Yang-Mills simulations. It will be important in future to quantify this comparison and understand better how BEC formation is influenced by $2\leftrightarrow 3$ processes at weak coupling.

We note that for self-interacting scalar theories, the existence of a BEC is firmly established, for the same expanding geometry as the Glasma~\cite{Berges:2014bba}. A big puzzle discussed in \cite{Berges:2014bba} was why the simulations of scalar theories showed identical universal behavior to that of the Glasma. A conventional kinetic description of a scalar theory would not capture its dynamics. In contrast, an effective kinetic description with \ttt scattering, albeit with couplings modified by infrared dynamics, reproduces the infrared behavior in numerical simulations in a controlled fashion~\cite{Orioli:2015dxa}. This is similar to our BEC+\ttt scattering effective description albeit it should be noted that our BEC will not capture important features of the IR dynamics in the scalar case. It is challenging to perform controlled computations of infrared dynamics in a gauge theory as have been performed for a scalar theory, or even interpret this IR dynamics definitively as a BEC. However, as we noted previously, recent computations of the spatial string tension in the Glasma~\cite{Mace:2016svc,Berges:2017igc} provide an essential guidepost for future studies.

The other key result of this paper was the first numerical simulation of the temporal evolution of quark distributions in the expanding overoccupied Glasma. We showed that light quark flavors obey identical scaling distributions to those of gluons. Quarks however take longer to achieve the scaling behavior because their effective interactions are weaker than those of gluons. The scaling results have already been employed to estimate the photon yield in the Glasma~\cite{Berges:2017eom}; our results there showed that photon yields from the Glasma are significant relative to those from the thermalized QGP. We also studied the rate at which chemical equilibration is achieved. This result is sensitive to the initial quark occupancies. For maximal initial quark occupancies, chemical equilibration is approached rapidly; it remains to be seen how inelastic collisions which become influential for times $\tau \gtsim Q_s^{-1}\,\alpha_s^{-3/2}$ modify this result. 

As an outlook for future work, it will be important to perform more detailed comparisons of our effective kinetic framework both with classical-statistical Yang-Mills simulations and kinetic simulations in the bottom-up framework that include inelastic scattering contributions. These comparisons are relevant both for our results for $N_f=0$ and at $N_f=3$. Further, a deeper theoretical understanding of our results, in the context of the puzzles outlined is desirable. A useful guide is the body of work employing two-particle irreducible $1/N$ techniques that have proved powerful in the context of scalar theories~\cite{Berges:2015kfa}. These also provide guidance in the extrapolation of weak coupling techniques to couplings relevant for heavy-ion collisions~\cite{Berges:2016nru}. Finally, we note that in our discussion of photon yields in \cite{Berges:2017eom}, we did not consider the contribution to photon production from the interaction of quarks with the BEC. Such a possibility was discussed previously in \cite{Chiu:2012ij}. It will be useful to reconsider the possible effects of a transient BEC in this context, and in other contexts~\cite{Begun:2015ifa,Gangadharan:2016vez,Osada:2017oxe}, such as for instance the intriguing results on Hanbury-Brown--Twiss correlations from the ALICE experiment at the LHC~\cite{Adam:2015pbc}.

\section*{Acknowledgements}
We would like to thank Juergen Berges, Kirill Boguslavski, Aleksi Kurkela and Alexander Rothkopf for valuable discussions and comments.
R.~V.~is supported under DOE Contract No.~DE-S{C0012}704. 
He would like to thank the Institut f\"{u}r Theoretische Physik, Universit\"{a}t Heidelberg for kind hospitality and support via the Excellence Initiative during the early stages of this work.
This work is part of and supported by the DFG Collaborative Research Centre ``SFB~1225~(ISOQUANT)".

\appendix
\section{Numerical method} \label{sec:num_method}
In this appendix, we will discuss details of the numerical method employed to solve the kinetic equations. 
Since the kinetic equation for gluons Eq.~\eqref{kine_g1}, and that for quarks Eq.~\eqref{kine_q1}, have the same structure, it is sufficient to discuss only Eq.~\eqref{kine_g1}. 
In terms of the variables $p=\sqrt{\pperp^2+p_z^2}$ and $\kappa =p_z/p$, Eq.~\eqref{kine_g1} can be rewritten as
\begin{align} \label{kine_g2}
&\left( \frac{\partial}{\partial \tilde{\tau}} -\frac{p\kappa^2}{\tilde{\tau}} \frac{\partial}{\partial p} 
-\frac{\kappa (1-\kappa^2)}{\tilde{\tau}} \frac{\partial}{\partial \kappa} \right) 
\tilde{f} ( \tilde{\tau} ,p, \kappa ) \notag \\
&\hspace{10pt}
= -\tilde{\mathcal{L}} \frac{1}{p^2} \frac{\partial}{\partial p} \mathcal{R} ( \tilde{\tau} ,p, \kappa ) 
+\tilde{\mathcal{L}} \tilde{I}_a \frac{1}{p^2} \left[ (1-\kappa^2 )\frac{\partial^2}{\partial \kappa^2}
-2\kappa \frac{\partial}{\partial \kappa} \right] \tilde{f} ( \tilde{\tau} ,p, \kappa ) 
+\frac{C_F N_f}{N_c^2 f_0} \tilde{S} \, ,
\end{align}
where $\mathcal{R}$ denotes the (rescaled) flow in the radial direction,
\begin{equation}
\mathcal{R} ( \tilde{\tau} ,p, \kappa ) 
= -\tilde{I}_a p^2 \frac{\partial}{\partial p} \tilde{f} ( \tilde{\tau} ,p, \kappa )
-\tilde{I}_b p^2 \tilde{f} ( \tilde{\tau} ,p, \kappa ) \left[ f_0^{-1}+\tilde{f} ( \tilde{\tau} ,p, \kappa )\right] \, .
\end{equation}
The form of this flow is exactly the same as that in earlier studies for isotropic systems, except that now $\mathcal{R}$ also depends on the angular variable  $\kappa$. 
The second term on the r.h.s of Eq.~\eqref{kine_g2} describes diffusion in $\kappa$--this term is of course absent in the isotropic case.

To solve the kinetic equations numerically, we discretize the momentum variables $p$ and $\kappa$ into $N_p$ and $N_\kappa$ grids, respectively, as
\begin{equation}
p_i = p_\text{min} e^{(i-1)\Delta u} \, , \hspace{20pt}
\Delta u = \frac{\log p_\text{max} -\log p_\text{min}}{N_p-1} \, , \hspace{20pt}
i=1,2, \cdots , N_p \, ,
\end{equation} 
and
\begin{equation}
\kappa_j = \sin \left[ (j-1)\Delta \theta \right] \, , \hspace{20pt}
\Delta \theta = \frac{\pi/2}{N_\kappa -1} \, , \hspace{20pt}
j=1,2, \cdots , N_\kappa \, . 
\end{equation}
Before discretizing Eq.~\eqref{kine_g2}, we rewrite $p\partial \tilde{f}/\partial p$ that appears on the l.h.s as
\begin{equation}
p\frac{\partial \tilde{f}}{\partial p} = \frac{1}{p^2} \frac{\partial (p^3 \tilde{f})}{\partial p} -3\tilde{f} \, .
\end{equation}
This ``integration by parts'' is helpful in preserving number conservation (Eq.~\eqref{n_conv}) to high numerical accuracy. 
We discretize Eq.~\eqref{kine_g2} as
\begin{align} \label{kine_g_disc}
\frac{\partial f_{ij}}{\partial \tilde{\tau}} 
&= -\frac{1}{p_i^2} \frac{\mathcal{R}_{ij} -\mathcal{R}_{i-1,j}}{\Delta p_i}
+\frac{\tilde{I}_a}{p_i^2} \left[ (1-\kappa_j^2 ) \left( \frac{\partial^2 f}{\partial \kappa^2} \right)_{\! ij}
-2\kappa_j \left( \frac{\partial f}{\partial \kappa} \right)_{\! ij} \right] \notag \\
&\hspace{10pt}
+\frac{\kappa_j^2}{\tilde{\tau}} \frac{1}{p_i^2} \left( \frac{\partial (p^3f)}{\partial p} \right)_{\! ij}
-3\frac{\tilde{\kappa}_j^2}{\tilde{\tau}} f_{ij} 
+\frac{\kappa_j -\kappa_j^3}{\tilde{\tau}} \left( \frac{\partial f}{\partial \kappa} \right)_{\! ij} 
+\frac{C_F N_f}{N_c^2 f_0} \tilde{S}_{ij} \, ,
\end{align}
where
\begin{equation}
f_{ij} = \tilde{f} (\tilde{\tau} ,p_i ,\kappa_j ) \, ,
\end{equation}
\begin{equation}
F_{ij} = F (\tilde{\tau} ,p_i ,\kappa_j ) \, ,
\end{equation}
\begin{equation} \label{dp}
\Delta p_i = p_i - p_{i-1} \hspace{10pt} (i=1,\cdots ,N_p) \, ,
\end{equation}
\begin{equation}
\Delta \kappa_j = \kappa_{j+1} -\kappa_j \hspace{10pt} (j=1,\cdots ,N_\kappa -1) \, ,
\end{equation}
\begin{equation}
\mathcal{F}_{ij} = -\tilde{I}_a p_i^2 \frac{f_{i+1,j}-f_{ij}}{\Delta p_{i+1}} -\tilde{I}_b p_i^2 f_{ij} \left( f_0^{-1} +f_{ij} \right) 
\hspace{10pt} (i=1,\cdots ,N_p-1); \hspace{10pt} \mathcal{F}_{N_p ,j} = 0 \, ,
\end{equation}
\begin{equation}
\tilde{S}_{ij} = \tilde{\mathcal{L}} \tilde{I}_c \frac{1}{p_i} \left[ F_{i,j} \left( f_0^{-1} +f_{ij} \right) -f_{ij} \left( 1-F_{ij} \right)\right] \, ,
\end{equation}
\begin{equation} \label{dp3f}
\left\{ \begin{array}{c}
\left( \frac{\partial (p^3f)}{\partial p} \right)_{\! ij} = \frac{p_{i+1}^3 f_{i+1,j} -p_{i-1}^3 f_{i-1,j}}{\Delta p_{i+1}+\Delta p_i} 
\hspace{10pt} (i=1,\cdots ,N_p-1); \\
\left( \frac{\partial (p^3f)}{\partial p} \right)_{\! N_p, j} = \frac{p_{N_p}^3 f_{N_p,j} -p_{N_p-1}^3 f_{N_p-1,j}}{\Delta p_{N_p}} \, ,
\end{array} \right. 
\end{equation}
\begin{equation}
\left( \frac{\partial f}{\partial \kappa} \right)_{\! ij} = \frac{f_{i,j+1}-f_{i,j-1}}{\Delta \kappa_j +\Delta \kappa_{j+1}} 
\hspace{10pt} (j=2,\cdots ,N_\kappa -1); \hspace{10pt}
\left( \frac{\partial f}{\partial \kappa} \right)_{\! iN_\kappa} = \frac{f_{i,N_\kappa}-f_{i,N_\kappa-1}}{\Delta \kappa_{N_\kappa-1}} \, ,
\end{equation}
\begin{equation}
\left( \frac{\partial^2 f}{\partial \kappa^2} \right)_{\! ij} 
= \frac{\frac{f_{i,j+1}-f_{i,j}}{\Delta \kappa_j} -\frac{f_{i,j}-f_{i,j-1}}{\Delta \kappa_{j-1}}}{\frac{1}{2} (\Delta \kappa_j +\Delta \kappa_{j+1})} 
\hspace{10pt} (j=2,\cdots ,N_\kappa -1); \hspace{10pt}
\left( \frac{\partial^2 f}{\partial \kappa^2} \right)_{\! i1} 
= 2\frac{f_{i,2}-f_{i,1}}{\Delta \kappa_1^2} \, ,
\end{equation}
and
\begin{equation}
\left\{ \begin{array}{c}
\tilde{\kappa}_j^2 = \frac{1}{3} \left( \kappa_{j+1}^2 +\kappa_{j+1} \kappa_{j-1} +\kappa_{j-1}^2 \right) 
\hspace{10pt} (j=2,\cdots ,N_\kappa -1); \\
\tilde{\kappa}_1^2 = \frac{1}{3} \left( \kappa_2^2 +\kappa_2 \kappa_1 +\kappa_1^2 \right) \, ; \hspace{10pt}
\tilde{\kappa}_{N_\kappa}^2 = \frac{1}{3} \left( \kappa_{N_\kappa}^2 +\kappa_{N_\kappa} \kappa_{N_\kappa-1} +\kappa_{N_\kappa-1}^2 \right) \, .
\end{array} \right. 
\end{equation}
In Eqs.~\eqref{dp} and \eqref{dp3f}, $p_0$ should be set to zero. For this choice in the discretization for the equation of motion,
the integrals $\tilde{I}_a$ and $\tilde{I}_b$ can be replaced by
\begin{align}
\tilde{I}_a &= \frac{1}{2\pi^2} \sum_{i=1}^{N_p-1} \Delta p_{i+1} \, p_i^2 \sum_{j, \, \text{trap}} \Delta \kappa_j \, 
\left[ f_{ij} \left( f_0^{-1} +f_{ij} \right) +\frac{N_f}{N_c f_0^2} F_{ij} \left( 1-F_{ij} \right) \right] \, , \\
\tilde{I}_b &= \frac{1}{2\pi^2} \sum_{i=1}^{N_p-1} \Delta p_i \, (p_i +p_{i-1}) \sum_{j, \, \text{trap}} \Delta \kappa_j \, 
\left[ f_{ij} +\frac{N_f}{N_c f_0} F_{ij} \right] \, , 
\end{align}
where $\sum_{j,\text{trap}}$ denotes the trapezoidal sum,
\begin{equation}
\sum_{j, \, \text{trap}} \Delta \kappa_j X_j = \sum_{j=1}^{N_\kappa-1} \Delta \kappa_j \frac{X_j+X_{j+1}}{2} \, .
\end{equation}
Note that the discretization of  $\int\! dp$ is different for $\tilde{I}_a$ and $\tilde{I}_b$. 
This is done in order to preserve energy conservation (Eq.~\eqref{ene_conv}) to high accuracy. 
For the discretization of $\tilde{I}_c$, there is no guidance from conservation laws.
For simplicity, we employ the same discretization as for $\tilde{I}_b$:
\begin{equation}
\tilde{I}_c = \frac{1}{4\pi^2} \sum_{i=1}^{N_p-1} \Delta p_i \, (p_i +p_{i-1}) \sum_{j, \, \text{trap}} \Delta \kappa_j \, 
\left[ f_{ij} +\frac{1}{f_0} F_{ij} \right] \, .
\end{equation}   
As noted, the kinetic equation for quarks, Eq.~\eqref{kine_q1}, is discretized exactly in the same way.  

To solve Eq.~\eqref{kine_g_disc}, a boundary condition for the radial flow $\mathcal{R}_{ij}$ at $i=0$ is necessary. 
Since we consider only  number conserving elastic collisions, the onset of the Bose-Einstein condensate occurs at a finite time for an overoccupied initial condition \cite{Blaizot:2013lga}. 
When the condensate is generated, special care is necessary for the boundary condition at $p=0$. 
We adopt the method presented in Ref.~\cite{Blaizot:2014jna}. 
Before the onset of the BEC, the gluon distribution is less singular than $1/p$. 
Therefore, $\lim_{p\to 0} \mathcal{R} (p,\kappa ) =0$. 
Once the BEC is formed, the distribution shows a $1/p$ behavior in the infrared, and
the flow into the deep infrared becomes nonzero:
\begin{equation}
\lim_{p\to 0} \mathcal{R} (p,\kappa ) = \tilde{I}_a c -\tilde{I}_b c^2 \, ,
\end{equation}
where $c$ is the coefficient of the $1/p$ term of the gluon distribution; $\tilde{f}(p)\simeq c/p$. 
In numerical computations, we employ the following boundary condition
\begin{equation} \label{flow0}
\mathcal{R}_{0j} = \begin{cases}
0 & (\tau <\tau_c ) \\
\tilde{I}_a c_j -\tilde{I}_b c_j^2 & (\tau >\tau_c )
\end{cases}
\end{equation}
where the coefficient $c_j$ is determined by
\begin{equation} \label{cj}
c_j = p_\text{min} f_{1j} \, ,
\end{equation}
and $\tau_c$ is the time for the onset of condensation, defined as the instant that $\tilde{I}_b c_j^2$ becomes larger than $\tilde{I}_a c_j$. 
In principle, $\tau_c$ can depend on $j$. 
However for sufficiently small $p_\text{min}$, it does not depend on $j$. 
For quarks, the radial flow at the origin is always vanishing because the distribution for fermions is bounded to be below unity. 
Therefore the boundary condition for the quark kinetic equation is $\mathcal{R}_{0j}=0$.  

For the time evolution, we employ the alternative direction implicit method.
However, we treat the nonlinear terms in $f$ and $F$ explicitly.

\section{Bose-Einstein condensate} \label{sec:BEC}
The Landau kinetic equation for the \ttt elastic collisions was employed previously to discuss the onset of Bose-Einstein condensation \cite{Blaizot:2013lga,Blaizot:2014jna}.
However number changing inelastic processes play a crucial role in the infrared region and they may hinder the onset of the condensate \cite{Kurkela:2011ti,York:2014wja,Blaizot:2016iir}. While they will ensure that any BEC  formed is transient in the ``long run", a BEC or more generally a large IR occupancy may play a role in the classical regime that is the focus of this study. For the expanding Glasma, a role of inelastic processes in BEC formation needs more careful study, for a range of couplings, but is outside the scope of the investigation here. 

Here we will focus on the numerical implementation of BEC formation in \ttt scattering and discuss its properties. 
Although an accurate description of condensate formation  and IR dynamics is outside of the range of applicability of our kinetic equations, BEC formation affects the applicability of our kinetic equations because of the boundary condition in momentum space, as discussed in the previous section. 
For simplicity, we will show only results for $N_f=0$ -- without quarks. 
As long as gluons are initially overoccupied, the effect of quarks on the gluon condensate is negligible. 
The effects of quarks for moderate values of $f_0$ in a non-expanding system have been discussed in Ref.~\cite{Blaizot:2014jna}. 

\begin{figure}[tb]
 \begin{center}
  \includegraphics[clip,width=8cm]{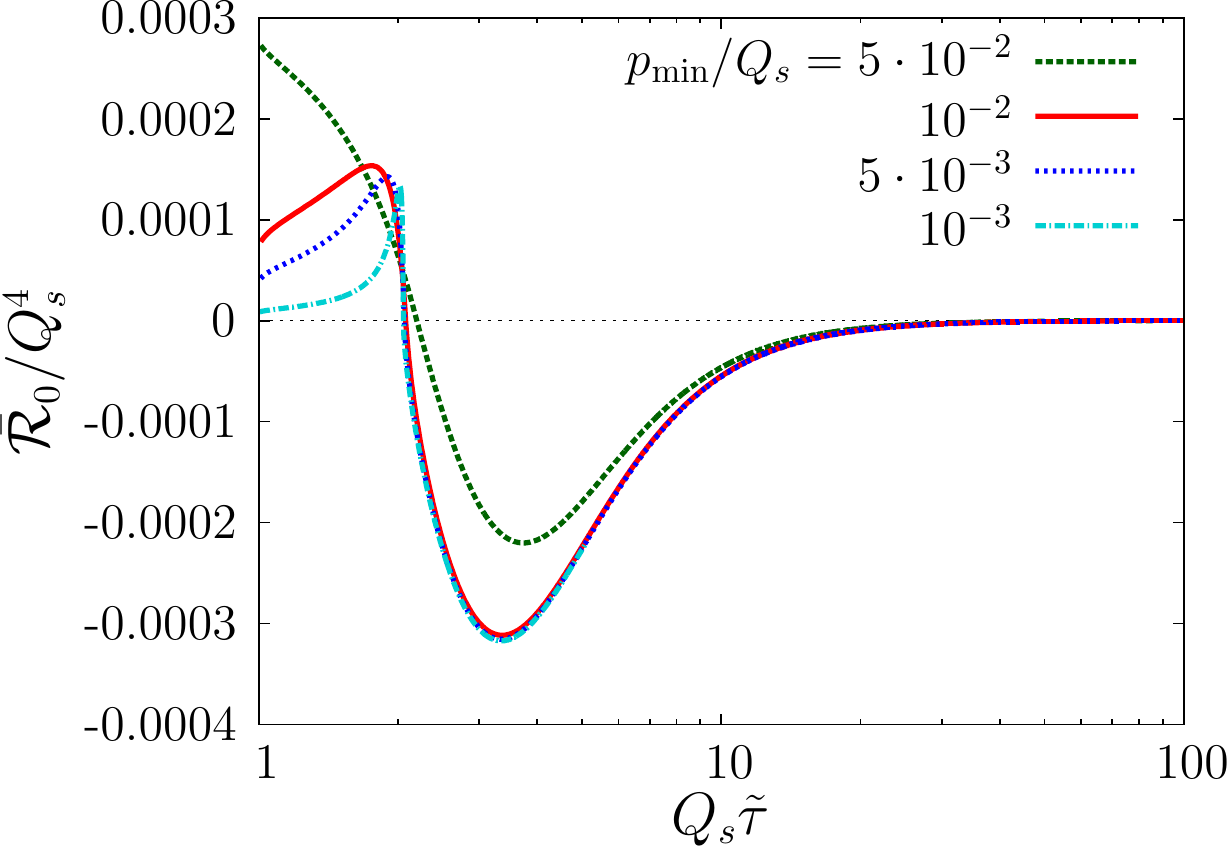} \vspace{-10pt}
  \caption{The infrared flow \eqref{av_flow} as a function of time. 
  Several values of the minimum momentum $p_\text{min}$ are compared. }
  \label{fig:flow}
 \end{center}
\end{figure}

As discussed in the previous section, the condensate forms when the flow to the deep infrared \eqref{flow0} sets in.
In Fig.~\ref{fig:flow}, we plot the flow, averaged over angle in momentum space,
\begin{equation} \label{av_flow}
\bar{\mathcal{R}}_0 = \frac{1}{N_\kappa} \sum_{j=1}^{N_\kappa} \left( I_a c_j -I_b c_j^2 \right) \, ,
\end{equation}
as a function of time. $c_j$ is computed by \eqref{cj}. 
Results with different $p_\text{min}$ are compared\footnote{For $p_\text{min}/Q_s=10^{-2}$, we have used $N_p=500$. For other values of $p_\text{min}$, we adjusted $N_p$ so that $\Delta u$ is the same.}. 
At early times, we observe a positive flow corresponding to a UV cascade from infrared to higher momenta.
This is an artificial result that occurs because we have nonzero $p_\text{min}$.
Indeed, for smaller  $p_\text{min}$, positive flow is suppressed.
This observation justifies the prescription \eqref{flow0}. 

After a time $Q_s \tilde{\tau}_c \simeq 2$, the flow becomes negative, indicating particle flow into the infrared region
$p<p_\text{min}$. 
For sufficiently small $p_\text{min}$ ($p_\text{min}/Q_s \ltsim 10^{-2}$), the time $\tilde{\tau}_c$ and the behavior of $\bar{\mathcal{R}}_0$ after $\tilde{\tau}=\tilde{\tau}_c$ are insensitive to the value of $p_\text{min}$. 

\begin{figure}[tb]
 \begin{center}
  \includegraphics[clip,width=7.5cm]{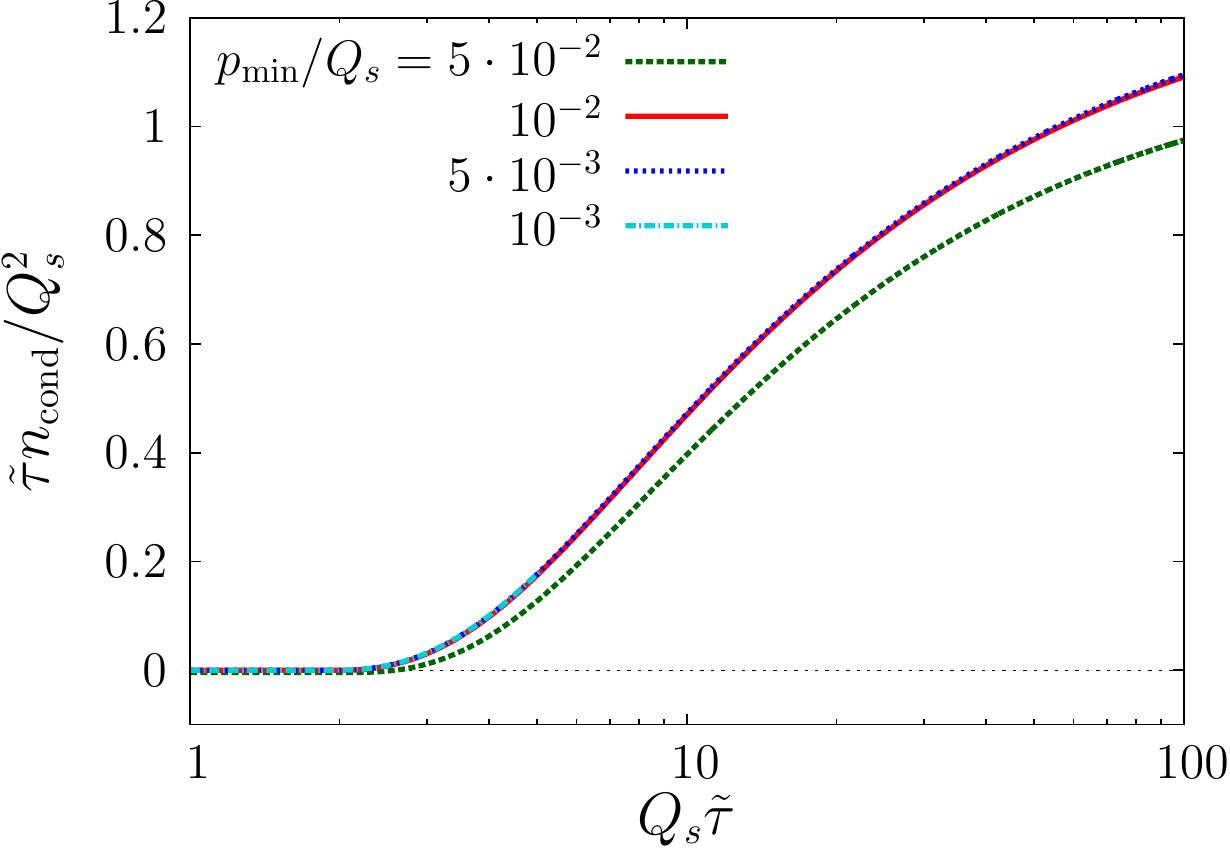} \vspace{-10pt}
  \caption{Time evolution of the particle number density in the BEC.
  Results for several values of the minimum momentum $p_\text{min}$ are compared. 
  The result is insensitive to the value of $p_\text{min}$ as long as it is sufficiently small.}
  \label{fig:ncond1}
 \end{center}
\end{figure}

Figure \ref{fig:ncond1} shows the time evolution of the BEC number density, defined as
\begin{equation}
\tilde{\tau} \tilde{n}_\text{cond} (\tilde{\tau}) = \tilde{\tau}_0 \tilde{n} (\tilde{\tau}_0) -\tilde{\tau} \tilde{n} (\tilde{\tau}) \, .
\end{equation}
Curves in the plot are insensitive to the value of $p_\text{min}$ for sufficiently small $p_\text{min}$.
Since $\tilde{\tau} \tilde{n}_\text{cond}$ can be regarded as the total particle number in the momentum sphere $p<p_\text{min}$, the insensitively to $p_\text{min}$ indicates that these particles condense to a zero mode. 

\begin{figure}[tb]
 \begin{center}
  \includegraphics[clip,width=8cm]{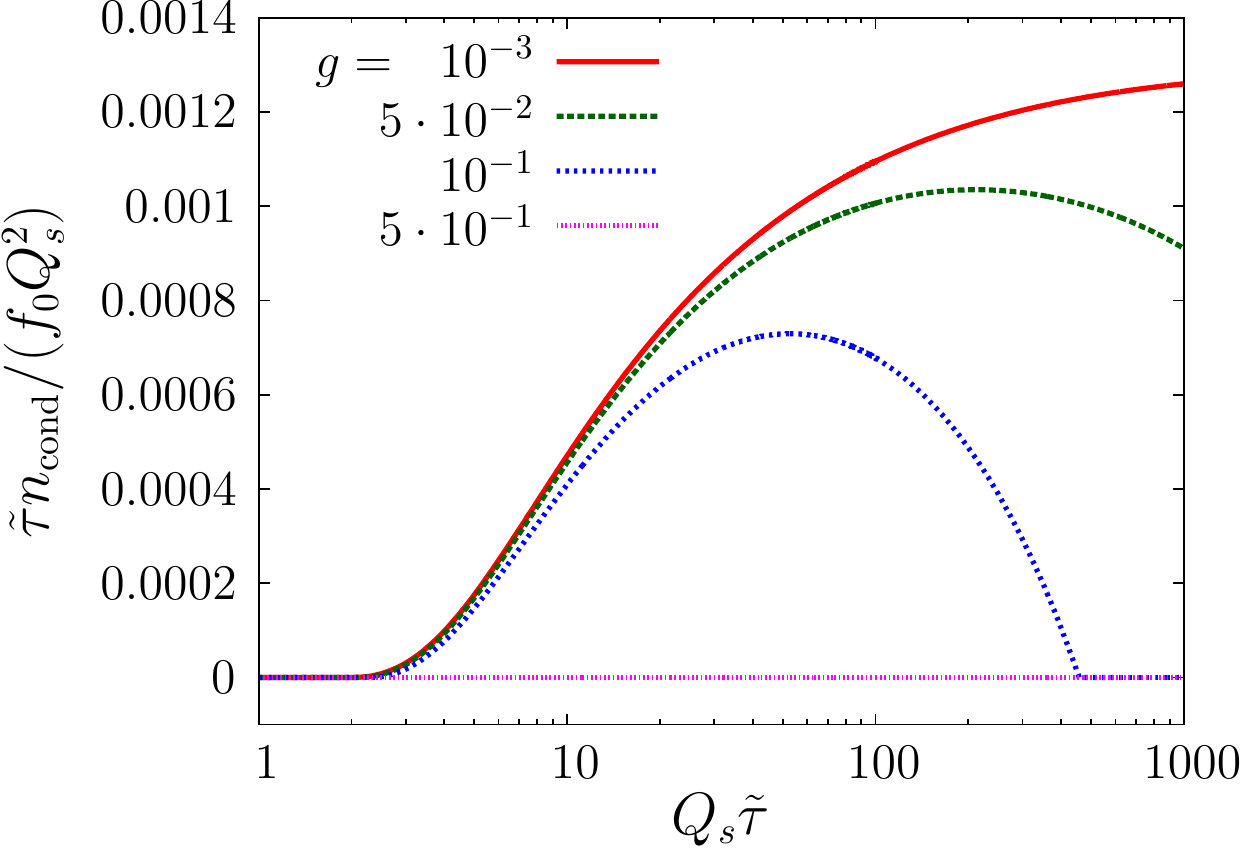} \vspace{-10pt}
  \caption{The number of particles in the BEC as a function of time for different values of the coupling.}
  \label{fig:ncond3}
 \end{center}
\end{figure}

The time evolution of particle number in the BEC is plotted in Fig.~\ref{fig:ncond3} for different values of the coupling.
The initial occupancy coefficient is fixed to be $n_0=0.1$. 
As we parametrize the initial occupancy to be $f_0=n_0/g^2$, different values of the coupling correspond to different initial occupation number. 
Other parameters used in these calculations are $\xi_0=1$ and $Q_s \tilde{\tau}_0=1$. 
In the figure, $\tilde{\tau} n$ divided by $f_0$ is plotted.
For $g=0.5$, which corresponds to $f_0=0.4$, $\tau n_\text{cond}$ stays zero indicating that the condensation does not occur.
For weaker coupling (larger initial occupancy),  condensation occurs around $Q_s (\tilde{\tau} -\tilde{\tau}_0)=1$
and $\tau n_\text{cond}$ begins increasing at earlier times.
In a nonexpanding system, the number of condensate particles approaches its equilibrium value, which can be nonzero for overoccupied initial conditions\footnote{One expects of course that 
quantum corrections at late times will destroy the BEC.}. 
In contrast,  condensation disappears in the expanding system, simply by its dilution from the expansion. 
Indeed, the increase in the condensate density stops at some point and a turnover is seen for $g=0.05$ and 0.1. 

\begin{figure}[tb]
 \begin{center}
  \includegraphics[clip,width=12cm]{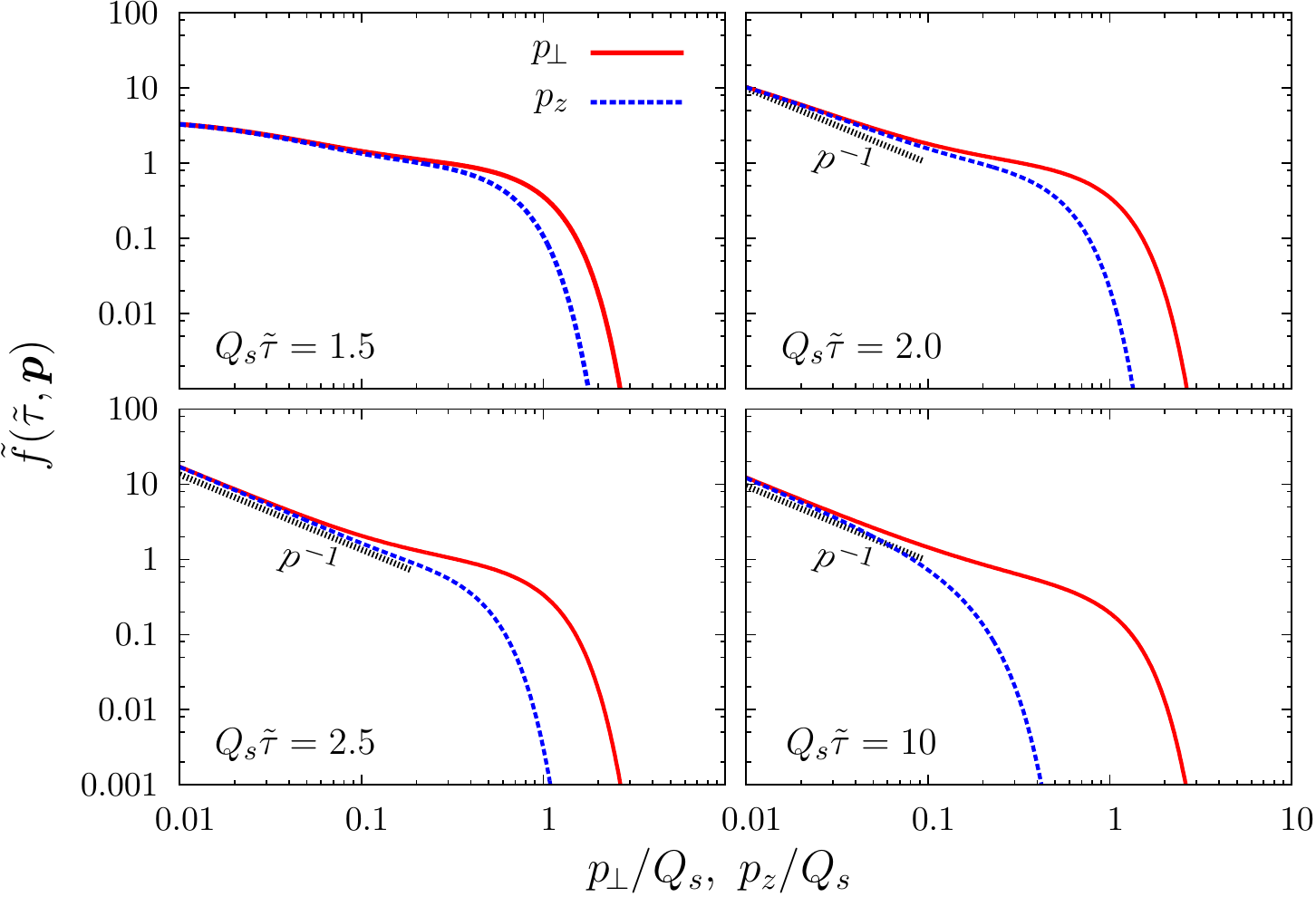} \vspace{-20pt}
 \end{center}
\caption{The transverse momentum distribution $\tilde{f}(\tilde{\tau}, \pperp, p_z=0)$ (red lines) and the longitudinal momentum distribution $\tilde{f}(\tilde{\tau}, \pperp=0, p_z)$ (blue dashed lines) at $Q_s \tilde{\tau}=1.5$, 2.0, 2.5, 10.
The time at which condensation occurs is $Q_s \tilde{\tau}_c \simeq 2.08$.}
\label{fig:dist_cond}
\end{figure}

We plot in Fig.~\ref{fig:dist_cond} the distribution function before and after the onset of the condensate.
The transverse momentum distribution evaluated at $p_z=0$ and the longitudinal momentum distribution at $\pperp=0$ are compared. 
The initial conditions $g=10^{-2}$, $n_0=0.1$ ($f_0=10^3$), $\xi_0=1$ and $Q_s \tilde{\tau}_0=1$ correspond to an overoccupied and isotropic distribution. 
The time for the onset of condensation is $Q_s \tilde{\tau}_c\simeq 2.08$. 
Although the distribution is initially isotropic, the hard sector $p\gtsim Q_s$ becomes quickly anisotropic due to the longitudinal expansion of the system. 
On the other hand, the soft sector $p\ltsim 0.1Q_s$ stays isotropic once the condensate is formed and maintains a $p^{-1}$ shape after 
$\tilde{\tau}=\tilde{\tau}_c$. However the region where the $p^{-1}$ shape is realized shrinks in the longitudinal direction due to the expansion.


\end{document}